\documentclass{emulateapj}
\shortauthors{Matthews et al.}
\shorttitle{\HI\ Observations of Cepheids}

\begin{document}
\newcommand{\ang}{\rm \AA}
\newcommand{\msun}{M$_\odot$}
\newcommand{\lsun}{L$_\odot$}
\newcommand{\days}{$d$}
\newcommand{\degree}{$^\circ$}
\newcommand{\ud}{{\rm d}}
\newcommand{\as}[2]{$#1''\,\hspace{-1.7mm}.\hspace{.0mm}#2$}
\newcommand{\am}[2]{$#1'\,\hspace{-1.7mm}.\hspace{.0mm}#2$}
\newcommand{\ad}[2]{$#1^{\circ}\,\hspace{-1.7mm}.\hspace{.0mm}#2$}
\newcommand{\lsim}{~\rlap{$<$}{\lower 1.0ex\hbox{$\sim$}}}
\newcommand{\gsim}{~\rlap{$>$}{\lower 1.0ex\hbox{$\sim$}}}
\newcommand{\HA}{H$\alpha$}
\newcommand{\HII}{\mbox{H\,{\sc ii}}}
\newcommand{\kms}{\mbox{km s$^{-1}$}}
\newcommand{\HI}{\mbox{H\,{\sc i}}}
\newcommand{\HeI}{\mbox{He\,{\sc i}}}
\newcommand{\jks}{Jy~km~s$^{-1}$}

\title{A Search for Mass Loss on the Cepheid Instability Strip using 
\HI\ 21-cm Line Observations}

\author{L. D. Matthews\altaffilmark{1}, M. Marengo\altaffilmark{2}, 
\& N. R. Evans\altaffilmark{3}}

\altaffiltext{1}{MIT Haystack Observatory, Off Route 40, Westford, MA
  01886 USA; lmatthew@haystack.mit.edu}
\altaffiltext{2}{Department of Physics \& Astronomy,
Iowa State University, Ames, IA 50011 USA}
\altaffiltext{3}{Harvard-Smithsonian Center for Astrophysics, 60
  Garden Street, MS-42, Cambridge, MA 02138 USA}

\begin{abstract}
We present the results of a search for \HI\ 21-cm line emission from the circumstellar
environments of four Galactic
Cepheids (RS~Pup, X~Cyg, $\zeta$~Gem, and T~Mon) based on observations
with the Karl G. 
Jansky Very Large Array. The observations were aimed at detecting
gas associated with previous or ongoing mass loss. Near the long-period Cepheid T~Mon,
we report the detection of a
partial shell-like structure  whose properties
appear consistent with originating from an earlier epoch of Cepheid mass
loss. At the distance of T~Mon,
the nebula would have a  mass (\HI+He) of
$\sim0.5M_{\odot}$, or $\sim$6\% of the stellar mass.   Assuming that one-third of the
nebular mass comprises swept-up interstellar gas,   we estimate an implied mass-loss rate of ${\dot
    M}\sim (0.6-2)\times10^{-5}~M_{\odot}$ yr$^{-1}$.
No clear signatures of circumstellar emission were found toward
$\zeta$~Gem, RS~Pup, or X~Cyg, although in each case, line-of-sight confusion
compromised portions of the spectral band. For the undetected stars, we derive model-dependent 
$3\sigma$ upper
limits on the mass-loss rates, averaged over their lifetimes on the instability strip, of 
$\lsim(0.3-6)\times10^{-6}~M_{\odot}$ yr$^{-1}$ and estimate the total
amount of mass lost to be less than a few per cent of the stellar
mass.   

\end{abstract}


\section{Introduction\protect\label{intro}}
\label{intro}
Cepheid variables serve as fundamental calibrators of the
cosmic distance scale, making these stars of vital importance
to extragalactic astronomy and cosmology
(Freedman et al. 2001; Di Benedetto 2013 and references therein).
However, 
important gaps remain in our understanding of the physics
and evolution of Cepheids.

One of the most confounding puzzles is the
decades-old problem known as the ``Cepheid mass
discrepancy": mass estimates based on stellar evolution
models are inconsistent with pulsation masses (derived
from the mass-dependent Period-Luminosity relation) and with masses inferred from
orbital dynamics (e.g., Christy 1968; Cox 1980; Pietrzy\'nski
et al. 2010). Discrepancies of
$\sim$10-20\% have persisted despite continued improvements in evolutionary
models (e.g., Bono et al. 2002; Caputo et al. 2005; Keller \& Wood
2006; Neilson et al. 2011). Proposed
solutions have included extra mixing, rotation, the need for better radiative
opacities, and perhaps most importantly, {\em mass-loss} (e.g., Cox 1980; Bono
et al. 2006;
Neilson et al. 2011, 2012a, b).

If mass loss is occurring during the Cepheid evolutionary phase, this could have important
implications for the use
of Cepheids as distance indicators, since the presence of
circumstellar material may add scatter to inferred
luminosities in the form of extra extinction in the visible and excess
emission at IR wavelengths (Neilson et al. 2009; 
Gallenne et al. 2013; Schmidt 2015). Indeed, accounting
for these effects may be key to resolving the discrepancy
between the Hubble constant determination from Cepheids compared with
that derived from Cosmic
Microwave Background measurements (e.g., Riess et al. 2016). 
Mass loss on the instability strip would also impact other
evolutionary stages of intermediate mass stars, 
including the relative lifetimes of the red and blue supergiant phases
(e.g., Dohm-Palmer \& Skillman 2002), and the determination of what
is the maximum initial mass of a star that will end its life as a white
dwarf rather than a supernova.

While Cepheid mass loss has been suspected for decades
(see review by Cox 1980), the direct and unambiguous detection of
escaped or outflowing material from Cepheids has 
proved to be challenging, leading to empirically estimated mass-loss
rates (or upper limits) spanning several orders of magnitude (${\dot M}\lsim10^{-12}$ to
$10^{-5}M_{\odot}$ yr$^{-1}$; McAlary \& Welch 1986; 
Welch \& Duric 1988; Deasy 1988; B\"ohm-Vitense \& Love 1994; Neilson
et al. 2009). 
However, a series of recent studies  has provided mounting evidence that not only is
mass loss common for stars on the instability strip, but it typically occurs at rates
high enough to significantly impact the star's evolutionary track.

In a study based on {\it Spitzer} infrared (IR) imaging data, Marengo et al. (2010b) reported the
discovery of a bow shock surrounding the Cepheid
archetype $\delta$~Cephei ($\delta$~Cep), providing direct evidence
for the existence of a stellar wind, and hence, ongoing mass loss at a
rate of $\sim10^{-7}~M_{\odot}$ yr$^{-1}$. 
Extended IR emission was also detected with {\it Spitzer} around several
other Cepheids by Barmby et al. (2011), including three stars with
extended emission seen in multiple IR bands and four other stars
with evidence for extended emission in at least one band.  In
addition, on
scales closer to the star, near- and mid-IR interferometry have revealed
what appear to be warm, dusty circumstellar envelopes on scales
ranging from a few stellar radii (M\'erand et al. 2006; Kervella et al. 2006; Gallenne et
al. 2013)
to several hundred AU (Kervella et al. 2009).

As noted by some authors (e.g., Schmidt 2015), observed IR
excesses and extended IR emission are not necessarily a {\em direct} product of ongoing
mass loss, particularly dusty mass loss. 
For example, in the case of $\delta$~Cep, the extended IR nebulosity
may be somehow linked with the presence of a binary companion
(Anderson et al. 2015), while in the case of RS~Pup, the vast circumstellar
nebulosity may represent a
pre-existing interstellar cloud (Kervella et al. 2009).
However, in both of these cases, there is evidence that a
stellar wind
has had a role in {\em shaping} the IR-emitting material. Similarly, 
Marengo et al. (2010a) suggested that near-IR emission seen close to
the star may result from shocked gas emission rather than
dust. Nonetheless, the presence of this emission is consistent with a
pulsationally-driven wind.

Another line of evidence for Cepheid mass loss comes from the work of
Neilson et al. (2012b), who analyzed the observed rates of period change,
${\dot P}$, for
a sample of 200 Galactic Cepheids and compared the results to stellar
evolution models. They found that models without mass loss could not
reproduce the observed ${\dot P}$ trends. However, mass loss on the Cepheid
instability strip at a mean rate
${\dot M}\sim 10^{-7}~M_{\odot}$ yr$^{-1}$ rectifies the models
with observations. For the specific
case of Polaris,  
Neilson et al. (2012a) concluded that a mass-loss rate
of ${\dot M}\sim 10^{-6}~M_{\odot}$ yr$^{-1}$ is necessary to account
for the secular period change of this star over the past $\sim$200
years.

Because of the moderate temperatures of Cepheids ($\sim$5000-6000~K), their winds are
expected to be predominantly neutral and atomic (Glassgold \& Huggins
1983), with at most, a modest ionized fraction (e.g., Engle et al. 2014). This
makes the \HI\ 21-cm line a potentially powerful tracer of Cepheid
outflows. Although contamination from interstellar \HI\
emission along 
the line-of-sight tends to be strong toward sources near
the Galactic plane, the finite
outflow velocity of the wind is expected in most cases to shift a
portion of the circumstellar gas outside
of the velocity range most strongly affected by line-of-sight emission. In addition,
interferometers act as spatial filters against the largest
scale components of the line-of-sight emission, which can aid in
disentangling circumstellar signals from foreground and/or background signals
(Bowers \& Knapp 1987; Matthews \&
Reid 2007; Le~Bertre et al. 2012). 

Motivated by these factors,
Matthews et al. (2012; hereafter M12) used the legacy Very Large Array to 
observe $\delta$~Cep in the \HI\ 21-cm line with the goal of searching for
a gaseous counterpart to the stellar wind revealed by the {\it
  Spitzer} observations of Marengo et al. (2010b).
Based on the \HI\ data, M12 reported the discovery of an extended
\HI\ nebula ($\sim13'$, or 1~pc across) surrounding the position of
$\delta$~Cep. This nebula exhibits a head-tail morphology, consistent with
debris that was ejected from the star and subsequently sculpted by its interaction
with the interstellar medium (ISM). M12 derived an outflow velocity for the wind of $V_{\rm
  o}\approx35.6\pm$1.2~\kms---the first ever
directly measured from a Cepheid---and constrained the mass-loss rate
to be ${\dot M}\approx (1.0\pm0.8)\times10^{-6}~M_{\odot}$ yr$^{-1}$.  

If similar \HI\ envelopes are present around 
other Cepheids, this would have profound implications for our
understanding of these stars and our ability to constrain their
mass-loss and evolutionary histories.
For this reason, we have undertaken \HI\ imaging observations of a
sample of four additional Galactic Cepheids using the upgraded Karl F. Jansky Very Large Array
(VLA) of the National Radio Astronomy Observatory\footnote{The National
Radio Astronomy Observatory is operated by Associated Universities,
Inc., under cooperative agreement with the National Science
Foundation.}. 
As described below,
these observations have uncovered evidence for circumstellar material
associated with one additional Cepheid and
allow us to place limits on the mass of circumstellar material
associated with the three remaining targets. 

\section{Target Selection\protect\label{targets}}
\label{targets}
A sample of four Galactic Cepheids was targeted in the present study: RS~Puppis
(RS~Pup), T~Monocerotis (T~Mon), X~Cygni (X~Cyg), and $\zeta$~Geminorum ($\zeta$~Gem).
Some of their properties are summarized in Table~1. 

The long-period Cepheid RS~Pup is one of the brightest known Cepheids in the
Galaxy. Based on its rate of period change it is thought to be on its
third crossing of the instability strip (Berdnikov et al. 2009). 
This star is unique among Galactic Cepheids in being 
surrounded by an extended ($\sim2'$ across) optical reflection nebula
(Westerlund 1961). This nebula is also visible  in
the IR (e.g., McAlary \& Welch 1986), including 
the multi-band {\it Spitzer} observations presented by Kervella et al. (2009) and
Barmby et
al. (2011). 
Although it was suggested early-on that the
RS~Pup nebula may be the result of mass loss (either during the
Cepheid phase or a previous red giant stage; Havlen 1972),  Kervella
et al. (2009, 2012) argued that the bulk of the nebula instead comprises
cold, dusty interstellar material (see also
Deasy 1988; Barmby et al. 2011) that was shaped and compressed by a
stellar wind or outflow, possibly during an earlier evolutionary phase
as a rapidly rotating B dwarf. However, Kervella et
al. (2009) also detected evidence of a warm emission component toward RS~Pup at
10~$\mu$m, with spatial
scales of $\sim$100 to 1000~AU. This latter component is interpreted
as a hallmark of ongoing Cepheid-phase mass loss (see
also Gallenne et al. 2011).

Among the remaining stars in the Barmby et al. (2011) sample that showed
either clear or possible extended emission, only two are far enough north to
observe with the VLA: T~Mon and X~Cyg. Both of
these stars show extended emission at 8.0~$\mu$m, but only tentative
detections at 24$\mu$m and/or 70$\mu$m. 

X~Cyg is one of the most luminous
classical Cepheids visible from
the northern hemisphere and is believed to be on its third crossing of
the instability strip (Turner 1999). Observations to date provide no compelling evidence for a 
companion (Evans 1984, 1992; Turner 1998), although a low-mass
companion with an orbit in the plane of the sky cannot be excluded.
T~Mon,
on the other hand, is a well-known binary (Mariska et
al. 1980; Coulson 1983) with an orbital period between 90
and 260 years (Evans et al. 1999). The companion is a hot, chemically
peculiar A star that is itself most likely a binary in a short-period
orbit. 
T~Mon was studied by Gallenne et
al. (2013) using mid-IR interferometry, and these authors detected a
mid-IR excess that they attributed to the presence of a CSE. 

In contrast to the other three stars in Table~1, our fourth target,
$\zeta$~Gem, does not exhibit
any extended IR emission in the study of Barmby et
al. (2011). This medium period Cepheid was
included in the present sample in part to test whether observable signatures of
mass loss are exclusive to stars with extended IR emission. In
addition, this star was predicted to have minimal line-of-sight \HI\ contamination
because of the low cirrus levels seen in the images of Barmby
et al. and the modest \HI\ brightness temperatures along this
direction seen in previous \HI\ survey data. Based on its declining
period (Berdnikov et al. 2000), $\zeta$~Gem is believed to be on its
second crossing of the instability strip (e.g., Turner et al. 2006).

\begin{figure*}
%\vspace{-1.5cm}
\centering
\scalebox{0.7}{\rotatebox{90}{\includegraphics{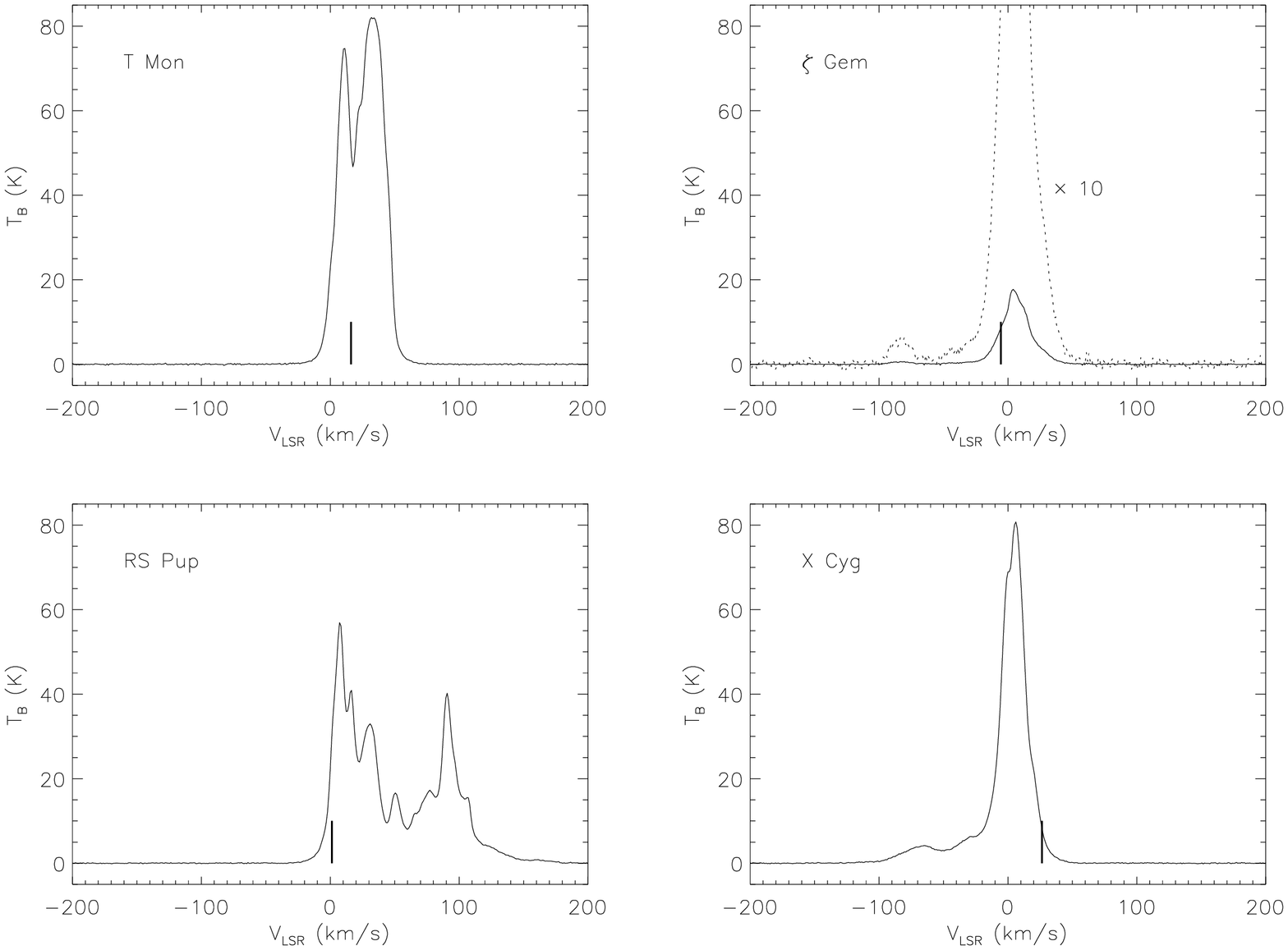}}}
\caption{Leiden/Argentine/Bonn (LAB) 
single-dish \HI\ spectra (Kalberla et al. 2005) toward each of the Cepheids
observed with the VLA, illustrating the line-of-sight interstellar
emission. 
Each spectrum shows brightness temperature in
  Kelvin plotted as a
  function of LSR velocity in \kms. 
For $\zeta$~Gem, the dotted line shows the original
  spectrum multiplied by a factor of 10. Vertical bars indicate the
  stellar systemic velocity.  }
\label{fig:LABspec}
\end{figure*}

To provide context for the interpretation of our VLA results 
and to illustrate the strength and velocity extent 
of the line-of-sight confusion toward each of our 
targets, we show in Figure~\ref{fig:LABspec} 
\HI\ spectra toward each stellar position (to within the
nearest \ad{0}{2}) extracted from the
Leiden Argentine Bonn (LAB) all-sky \HI\ survey (Kalberla et
al. 2005). Because the LAB spectra were obtained with single-dish
telescopes, they are not affected by filtering of large-scale
emission (i.e., missing short spacing information) and directly
measure the total beam-integrated \HI\ flux.  Note, however, that
the noise level in the LAB spectra ($\sim$0.6~Jy RMS) 
precludes the detection of typical weak circumstellar
signals, even outside of the velocity ranges affected by
interstellar contamination.

% Fig. 1
% 

%%%%%%%%%%%%%%%%%%%%%%%%%%%%%%%%%%%%%%%%%%%%%%%%%%%%%%%%%%%%%%%%%%%
% --- Table 1: 
\begin{deluxetable*}{lllcccccccccc}
\tabletypesize{\tiny}
\tablewidth{0pc}
\tablenum{1}
\tablecaption{Properties of the Target Stars}
\tablehead{
\colhead{Name} & \colhead{$\alpha$(J2000.0)} &
\colhead{$\delta$(J2000.0)} & \colhead{$l$} & \colhead{$b$}
 & \colhead{$V_{\star,\rm LSR}$} &  \colhead{$d$}  &
\colhead{P}& \colhead{$M$} & \colhead{$R_{\star}$} &
\colhead{$V_{\rm 3D}$} & \colhead{PA} & \colhead{Known}
  \\
\colhead{}     & \colhead{}     & \colhead{} & \colhead{(deg)} & \colhead{(deg)} &
\colhead{(\kms)} & \colhead{(pc)}   & \colhead{(days)} &
\colhead{($M_{\odot}$)} & \colhead{($R_{\odot}$)} &
\colhead{(\kms)} & \colhead{(degrees)} & \colhead{binary?}\\
  \colhead{(1)} & \colhead{(2)} & \colhead{(3)} &
\colhead{(4)} & \colhead{(5)} & \colhead{(6)} & \colhead{(7)}
& \colhead{(8)} & \colhead{(9)} & \colhead{(10)} & \colhead{(11)} & \colhead{(12)}
& \colhead{(13)}}

\startdata

T~Mon &  06 25 13.0 & 07 05 08.6 & 203.6 & $-2.6$& 16.1 &
1416 &  27.025 & 9.1 & 150 & 21.0 & 110 & Yes\\

$\zeta$~Gem & 07 04 06.5 & 20 34 13.1 & 195.8 & 11.9& $-5.47$ &
383  & 10.151 &6.4 & 73 &  14.9
& 326 & No\\

RS~Pup &  08 13 04.2  & $-$34 34 42.7 & 252.4 & $-0.2$& 1.28 &
1543 & 41.388 & 9.9 & 214 &  20.8 & 318 & No\\

X~Cyg & 20 43 24.2& +35 35 16.1 & 76.8& $-$4.3 & 26.42 & 981 &
16.386 & 7.5 & 105 & 52.5 & 248 & No\\

\enddata

\tablecomments{Units of right ascension are hours, minutes, and
seconds. Units of declination are degrees, arcminutes, and
arcseconds. Explanation of columns: (1) star name; (2) \& (3) right
ascension and declination (J2000.0); (4) \& (5) Galactic coordinates;
(6) systemic velocity relative to the Local Standard of
Rest (LSR); (7) adopted distance in parsecs; (8) 
pulsation period in days; (9) mass, computed as
log~$(M/M_{\odot})=0.297({\rm log}L_{\star}/L_{\odot})-0.259$ where
$L_{\star}$ is the stellar luminosity (Evans et al. 2013); (10)
mean stellar radius, computed from the period-radius relation of
Kervella et al. (2004); (11) space velocity,
computed following Johnson \& Soderblom (1987), adopting the solar
constants from Sch\"onrich et al. (2010) and the proper motions
from van Leeuwen (2007); (12)
position angle of space motion in the plane of the sky; (13) binary
status. Coordinates
were taken from
SIMBAD (http://simbad.harvard.edu). Distances and pulsation periods
were taken from Fernie et al. (1995).  The luminosities used to
compute the masses in column~9 are derived from
$M_{V,\star}-M_{V,\odot}+{\rm BC}=-2.5{\rm log}(L_{\star}/L_{\odot})$,
  where the solar absolute $V$ magnitude is $M_{V,\odot}$=4.73, the
  stellar absolute $V$ magnitude is taken to be $M_{V,\star}=-4.04-2.43({\rm log}P
  -1.0)$ (Evans et al. 2013), and the
  bolometric corrections (BC) are from Flower (1996).
 }

\end{deluxetable*}

%%%%%%%%%%%%%%%%%%%%%%%%%%%%%%%%%%%%%%%%%%%%%%%%%%%%%%%%%%%%%%%%%%%

%%%%%%%%%%%%%%%%%%%%%%%%%%%%%%%%%%%%%%%%%%%%%%%%%%%%%%%%%%%%%%%%%%%
% --- Table 2: Summary of observations
%
\begin{deluxetable*}{llccccc}
\tabletypesize{\tiny}
\tablewidth{0pc}
\tablenum{2}
\tablecaption{Summary of VLA Observations}
\tablehead{
\colhead{Star} & \colhead{Obs. date} & \colhead{$V_{\rm cent}$}  &
\colhead{$V_{\rm min}$,$V_{\rm max}$} &
\colhead{No.} & \colhead{$t$} & \colhead{Config.}\\
\colhead{}     &  \colhead{} & \colhead{(\kms)}
& \colhead{(\kms)} & \colhead{antennas}
& \colhead{(hours)}  \\
\colhead{(1)} & \colhead{(2)} & \colhead{(3)} & \colhead{(4)} & \colhead{(5)}
& \colhead{(6)} & \colhead{(7)}}

\startdata

$\zeta$~Gem & 2011-Dec-03 & $-5.5$& $-215.7$,205.6 & 27 & 2.75 & D \\

$\zeta$~Gem & 2011-Dec-04 & $-5.5$& $-215.7$,205.6 & 27 & 2.75 & D \\

X~Cyg & 2011-Dec-04 & 26.4 & $-183.8$,237.5 & 26 & 2.94 & D\\

X~Cyg & 2011-Dec-05 & 26.4 & $-183.8$,237.5 & 26 & 2.93 & D\\

T~Mon & 2011-Dec-05 & 15.6 & $-194.7$,226.6 & 25 & 2.86 & D\\

RS Pup& 2012-Jan-16 & 1.3 & $-209.0$,212.3 & 25 & 1.90 & DnC\\

RS Pup& 2012-Jan-17 & 1.3 & $-209.0$,212.3 & 25 & 1.90 & DnC\\

RS Pup& 2012-Jan-21 & 1.3 & $-209.0$,212.3 & 24 & 1.90 & DnC \\

\enddata

\tablecomments{Explanation of columns: (1) star name; (2) date of observation;
(3) LSR velocity at band  center; (4) minimum and maximum
LSR velocity covered by the observing band; (5) number of available antennas; (6)
total on-source integration time; (7) antenna configuration.}

\end{deluxetable*}

%%%%%%%%%%%%%%%%%%%%%%%%%%%%%%%%%%%%%%%%%%%%%%%%%%%%%%%%%%%%%%%%%%%

%%%%%%%%%%%%%%%%%%%%%%%%%%%%%%%%%%%%%%%%%%%%%%%%%%%%%%%%%%%%%%%%%%%
% --- Table 3: calibration
%
\begin{deluxetable*}{lccccl}
\tabletypesize{\tiny}
\tablewidth{0pc}
\tablenum{3}
\tablecaption{Calibration Sources}
\tablehead{
\colhead{Source} & \colhead{$\alpha$(J2000.0)} &
\colhead{$\delta$(J2000.0)} & \colhead{Flux Density (Jy)} &
\colhead{$\nu_{0}$ (GHz)} & \colhead{Date}
}

\startdata

3C48$^{a}$  & 01 37 41.2994 & +33 09 35.133 & 16.2632$^{*}$ & 1.4194  & 2011-Dec-04 \& 05\\
            & ...           & ...           & 16.2627$^{*}$ & 1.4193  & 2011-Dec-05\\

J0632+1022$^{b}$ & 06 32 15.3269 & 10 22 01.676 & 2.377$\pm$0.010 &
1.4198 & 2011-Dec-05\\

J0738+1742$^{c}$ & 07 38 07.3937 & 17 42 18.998 & 0.962$\pm$0.009 & 1.4204 & 2011-Dec-03\\
                 & ...           & ...          & 0.970$\pm$0.018 & 1.4204 & 2011-Dec-04\\

J0828-3731$^{d}$&08 28 04.7803 & $-$37 31 06.281 & 1.894$\pm$0.008 & 1.4203 & 2012-Jan-16\\
                & ...          & ...             & 1.898$\pm$0.010 & 1.4203 & 2012-Jan-17\\
                & ...          & ...             & 1.885$\pm$0.011 & 1.4203 & 2012-Jan-21\\

3C286$^{e}$ & 13 31 08.2879 & +30 30 32.958 & 15.0515$^{*}$ & 1.4196 & 2011-Dec-03\\
            & ...           & ...           & 15.0515$^{*}$ & 1.4196 & 2011-Dec-04\\
            & ...           & ...           & 15.0520$^{*}$ & 1.4195  & 2012-Jan-16\\
            &      ...      &    ...        & 15.0520$^{*}$ & 1.4195 & 2012-Jan-17\\ 
            &      ...      &        ...    & 15.0521$^{*}$ & 1.4195 & 2012-Jan-21\\ 

J2052+3635$^{f}$ & 20 52 52.0550 & +36 35 35.300  & 4.913$\pm$0.028 &
               1.4205 & 2011-Dec-04 \& 05\\

\enddata

\tablecomments{Units of right ascension are hours, minutes, and
seconds, and units of declination are degrees, arcminutes, and
arcseconds. $\nu_{0}$ is the frequency at which the flux density in
the fourth column was computed.}
\tablenotetext{*}{Flux densities were determined using the
  time-dependent coefficients from Perley \& Butler (2013). For 3C48, the flux
density $S_{\nu}$ as a function of frequency was taken to be
${\rm log}(S_{\nu}) = 1.3322 - 0.7688({\rm
  log}(\nu)) - 0.1952({\rm log}(\nu))^{2} +0.0593({\rm log}(\nu))^3$, where
  $\nu_{\rm GHz}$ is the frequency expressed in GHz. For 3C286, ${\rm
  log}(S_{\nu}) = 1.2515 - 0.4605({\rm
  log}(\nu)) - 0.1715({\rm log}(\nu))^{2} +0.0336({\rm
  log}(\nu))^3$.}

\tablenotetext{a}{Primary flux calibrator and bandpass calibrator for
 X~Cyg and T~Mon}
\tablenotetext{b}{Secondary gain calibrator for T~Mon}
\tablenotetext{c}{Secondary gain calibrator for
 $\zeta$~Gem}
\tablenotetext{d}{Secondary gain calibrator for
 RS~Pup}
\tablenotetext{e}{Primary flux calibrator and bandpass calibrator for
 RS~Pup and $\zeta$~Gem}
\tablenotetext{f}{Secondary gain calibrator for X~Cyg}

\end{deluxetable*}

%%%%%%%%%%%%%%%%%%%%%%%%%%%%%%%%%%%%%%%%%%%%%%%%%%%%%%%%%%%%%%%%%%%%

\section{Observations and Data Reduction\protect\label{observations}}
\label{observations}
\HI\ 21-cm line 
observations of each of our target stars were carried out with the
VLA in late 2011 and early 2012 (Table~2). 
To obtain maximum sensitivity to extended emission, the
most compact (D) configuration was used for X~Cyg, T~Mon, and
$\zeta$~Gem (with baselines ranging from
0.035-1.03~km) and the hybrid DnC configuration (0.035-1.6~km
baselines) was used for the
southern source RS~Pup. These configurations provide
sensitivity to emission on angular scales of up to $\sim$16$'$. The
primary beam of the VLA at our observing
frequency is $\sim 31'$. During each observing session, observations
of the target star were interspersed with observations of a
neighboring bright point source to provide calibration of the complex
gains. Additionally, either 3C48 or 3C286 was observed
as an absolute flux density calibrator and bandpass calibrator (see Table~2).

The WIDAR correlator was configured with 8 subbands across each of two
basebands,
both of which measured dual circular
polarizations. Because the two basebands sample the same data stream,
averaging them does not improve the RMS noise, hence
only data from the first baseband pair (A0/C0) were
used for the present analysis. Each  of the subbands had a bandwidth of
0.25~MHz with 128 spectral channels, providing a channel spacing of
1.95~kHz ($\sim$0.41~\kms). The 8 subbands were tuned to
contiguously cover a total
bandwidth of 2~MHz.

The data for each target star were taken with the central baseband frequency
tuned to approximately match the LSR velocity of the star (see
Tables~1 and 2). However, additional
observations of the phase and bandpass
calibrators were made with the frequency center shifted by $\sim\pm1.5$
to 2~MHz,
respectively (see Table~3). The offsets adopted for each case were
determined based on the velocity distribution of the 
Galactic \HI\ in the neighborhood of the star, as determined using
spectra from Kalberla et al. (2005). This approach mitigated contamination from Galactic
\HI\ emission in the band and thus permitted a more robust bandpass
calibration and more accurate bootstrapping of
the flux density scale.

All data processing was performed using the Astronomical Image Processing
System (AIPS; Greisen 2003). Data were loaded into AIPS from archival
science data model (ASDM) format
files using the {\sc BDFI}n
program from the Obit software package (Cotton 2008). This
step enables the creation of
tables containing on-line flags and system power measurements. Data
for RS~Pup were taken with 5-second integrations times, while data for
the other three stars were
recorded with 1-second integration times, but averaged to 5-second
time resolution in post-processing, prior to beginning the
calibration. 

After updating the antenna positions to the best available values and
flagging obviously corrupted data, an initial calibration of the
visibility data  was performed using the AIPS task {\sc
  TYAPL}, which uses the system power measurements to compute
data weights (Perley 2010).
Calibration of the bandpass and
the frequency-independent portion of the complex gains was
subsequently performed
using standard techniques, taking into account the special
considerations for recent VLA data
detailed in Appendix~E of the
AIPS Cookbook.\footnote{http://www.aips.nrao.edu/cook.html} In
addition, the gain
solutions for subbands affected by contamination from line emission
were interpolated from the adjacent
subbands when necessary.
Following these steps, time-dependent frequency
shifts were applied to the data to compensate for the Earth's motion,
and the data were Hanning smoothed in frequency, dropping every other spectral
channel. The resulting velocity resolution is $\sim$0.82~\kms.
For stars observed during multiple non-contiguous sessions, the data from the
different sessions were concatenated following this step.

%%%%%%%%%%%%%%%%%%%%%%%%%%%%%%%%%%%%%%%%%%%%%%%%%%%%%%%%%%%%%%%%%%%%

% --- Table 4: Deconvolved image characteristics
%
\begin{deluxetable*}{lccccrlcc}
\tabletypesize{\tiny}
\tablewidth{0pc}
\tablenum{4}
\tablecaption{Deconvolved Image Characteristics}
\tablehead{
\colhead{Source} & \colhead{Type} & \colhead{{$\cal R$}} & \colhead{Taper} &
\colhead{$\theta_{\rm FWHM}$} & \colhead{PA} &
\colhead{$\sigma_{\rm RMS}$} & \colhead{Continuum channels} & \colhead{Clean Boxes?}\\
\colhead{}  & \colhead{}   & \colhead{}  & \colhead{(k$\lambda$,k$\lambda$)}
& \colhead{($''\times ''$)} & \colhead{(degrees)} & \colhead{(mJy
beam$^{-1}$)} & \colhead{} & \colhead{}\\
\colhead{(1)} & \colhead{(2)} & \colhead{(3)} &
\colhead{(4)} & \colhead{(5)} &
\colhead{(6)} & \colhead{(7)} & \colhead{(8)} & \colhead{(9)}
}

\startdata

X~Cyg & cont. & 1 & ... &  \as{52}{2}$\times$\as{49}{8} & $-78.4$ &
0.19 & 8-228; 411-500 & Yes\\

X~Cyg & line & 5 & ... &  \as{58}{8}$\times$\as{55}{7} & $-70.6$ &
1.2 & 8-228; 411-500 & No\\

X~Cyg & line & 5 & 2,2 &  \as{96}{8}$\times$\as{93}{0} & $-39.6$ &
1.4 & 8-228; 411-500 & No\\

RS Pup & cont. & 1 & ... &  \as{49}{0}$\times$\as{40}{8} & 27.2 &
0.14 & 1-45; 290-512 & Yes\\

RS Pup & line & 5 & ... &  \as{55}{3}$\times$\as{44}{4} & 27.2 &
1.4 & 1-45; 290-512 & No\\ 

RS Pup & line & 5 & 4,4 &  \as{68}{7}$\times$\as{55}{2} & 13.7 &
1.5 & 1-45; 290-512 & No\\ 

$\zeta$ Gem & cont. & 1 & ... & \as{48}{8}$\times$\as{45}{4} & $-9.8$
& 0.10 & 10-196; 280-230; 370-502 & Yes\\

$\zeta$ Gem & line & 5 & ... & \as{55}{3}$\times$\as{50}{2} & $-14.1$
& 1.1 & 10-196; 280-230; 370-502 & No\\

$\zeta$ Gem & line & 5 & 2,2 & \as{96}{5}$\times$\as{87}{5} & $-24.6$
& 1.3 & 10-196; 280-230; 370-502 & No\\

T~Mon & cont. & 1 & ... & \as{60}{0}$\times$\as{45}{6} & $-3.1$ &
0.15& 1-185; 330-512 & Yes\\

T~Mon & line & 5 & ... & \as{69}{8}$\times$\as{48}{8} & $-2.5$ & 1.5&
10-185; 330-500 & Yes\\

T~Mon & line & 5 & 2,2 & \as{109}{3}$\times$\as{85}{5} & $-10.1$ & 1.8&
10-185; 330-500 & Yes\\

\enddata

\tablecomments{Explanation of columns: (1) target name; (2) indication
  of whether the image contains line or continuum emission; the
  continuum images comprise a single spectral channel representing an
  average of the line-free portions of the band; (3) AIPS robust
parameter used in image deconvolution; $\cal R$=+5 is comparable to
natural weighting; (4) Gaussian taper applied in $u$ and
$v$ directions, expressed as
distance to 30\% point of Gaussian in units of kilolambda;
(5) FWHM dimensions of
synthesized beam; (6) position angle of synthesized beam (measured
east from north); (7) RMS
noise per channel (1$\sigma$); (8) spectral channels used for continuum
subtraction (line data) or that were averaged to compute a continuum image;
(9) indication of whether or not clean boxes were used during
image deconvolution. }

\end{deluxetable*}

%%%%%%%%%%%%%%%%%%%%%%%%%%%%%%%%%%%%%%%%%%%%%%%%%%%%%%%%%%%%%%%%%%%

Prior to imaging the line data, continuum emission in the field was
subtracted using either a linear fit to the 
real and imaginary
components of the visibilities via the AIPS task {\small\sc UVLIN}, and/or
subtraction of a clean component model of the continuum (via the
AIPS task
{\small\sc UVSUB}). 
The portions of the band that were
determined to be line-free and were used to define the continuum are summarized in Column~7 of Table~4. 

An image of the 21-cm continuum
emission in the field of each target star was produced from the line-free portion
of the
band using the Clean deconvolution algorithm as implemented
in the AIPS task {\sc IMAGR} 
(see Table~4). The peak continuum flux densities measured within the primary
beam for each of the target fields were as follows: 15.6~mJy ($\zeta$ Gem);
30.2~mJy (T~Mon); 53.4~mJy (RS~Pup); 15.1~mJy (X~Cyg).
No continuum emission was detected coincident with any of the target
stars.

Deconvolved image cubes of the \HI\ line emission were also produced
using {\sc IMAGR}. 
For each target, data cubes were produced using various 
weighting  schemes for the visibilities. The characteristics of the
data cubes used for the present
analysis are summarized in Table~4. 

%-----------------------------------------------------------------------
\section{Observational Results\protect\label{results}}
Consistent with the single-dish spectra shown in 
Figure~\ref{fig:LABspec}, the interpretation of the \HI\
data for all of our Cepheid targets is impacted over portions of the
observing band by the presence of strong interstellar \HI\
emission along the line-of-sight. In general, this emission contains large-scale components
($>15'$) that
are poorly sampled by the VLA D configuration. 
Consequently, the true spatial structure of the gas over the affected velocity ranges
cannot be fully reconstructed in our
deconvolved images, 
leading to characteristic artifacts, including patterns of strong
positive and negative components that fill the field-of-view 
(cf. Figure~1 of M12 and Figure~\ref{fig:TMoncmapsred}
discussed below).  In the discussion that follows, we designate a spectral
channel as likely to be contaminated by Galactic emission if its noise
characteristics in a deconvolved data cube exhibit an excess of both
positive and negative pixels at a
significance of $>3\sigma$ compared with what is
expected from a Gaussian noise distribution. Because of the spatial
filtering effects of an interferometer, not all of the velocities containing line-of-sight
emission in the single-dish spectra in Figure~\ref{fig:LABspec} 
contain discernible contamination in the VLA maps, but in general, the
single-dish spectra are a good predictor of which velocity ranges will
be impacted.

% Fig. 2
% 
\begin{figure*}
%\vspace{-2.5cm}
\centering
\scalebox{0.9}{\rotatebox{0}{\includegraphics{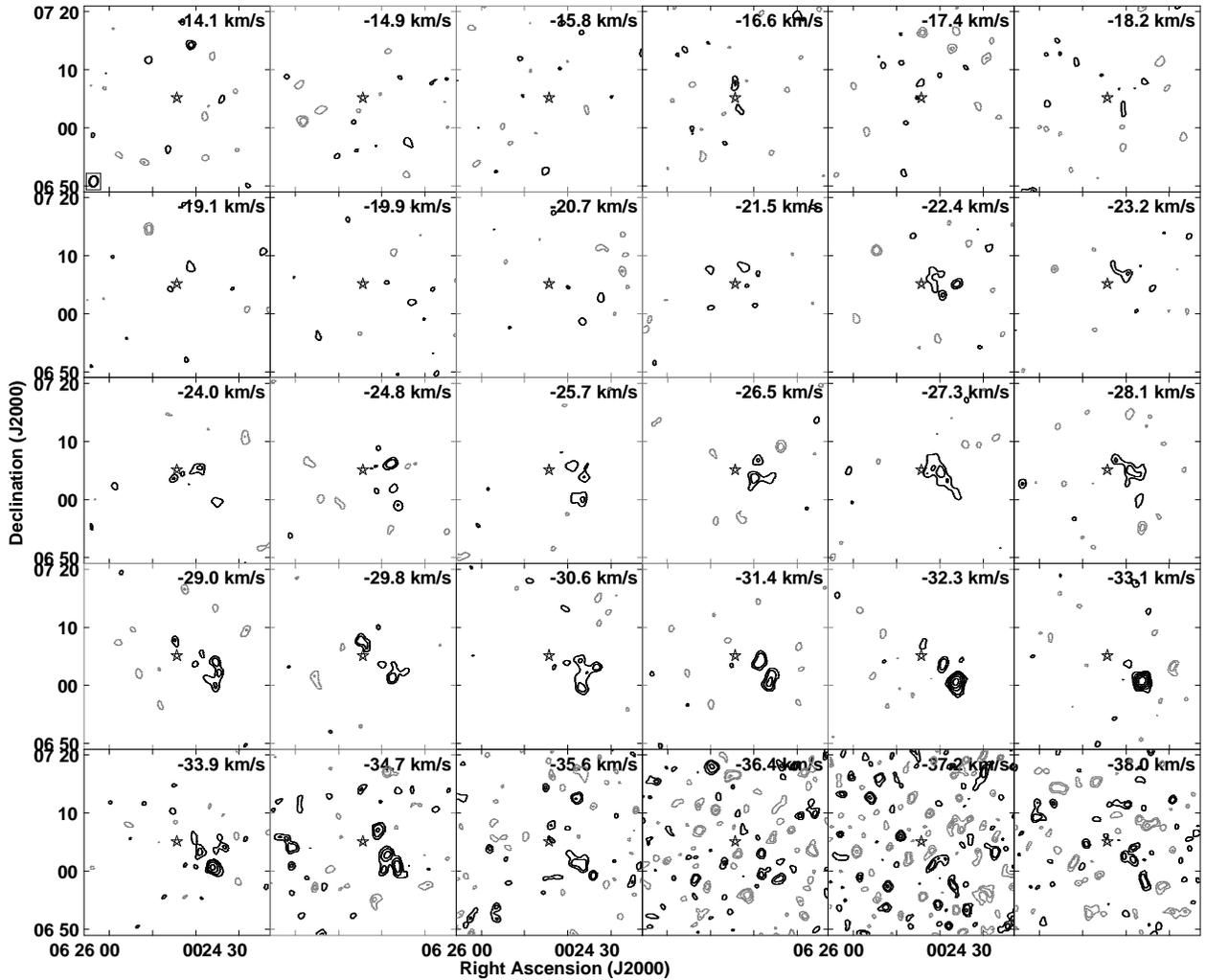}}}
\caption{\HI\ channel maps toward T~Mon, derived from a tapered
  version of the VLA data. The channels shown are blueshifted from
  the stellar systemic velocity of $V_{\star, {\rm LSR}}=16.1$~\kms.
A star symbol marks the stellar position. Contour levels are
  $-12,-8.5, -6, -4.2, -3,3,...12)\times1.8$ mJy beam$^{-1}$, where the lowest
  contour level is $\sim3\sigma$. Negative contours are shown as grey
  dotted lines;
  positive contours are plotted in black. The higher RMS noise in
  channels between $-24.7$ and $-38.8$~\kms\ results from aliased noise at
  the subband edges in the WIDAR correlator. The size of the
  synthesized beam ($109''\times85''$) is indicated in the lower left
  corner of the first panel.  }
\label{fig:TMoncmapsblue}
\end{figure*}

% Fig. 3
% 
\begin{figure*}
%\vspace{-1.5cm}
\centering
\scalebox{0.9}{\rotatebox{0}{\includegraphics{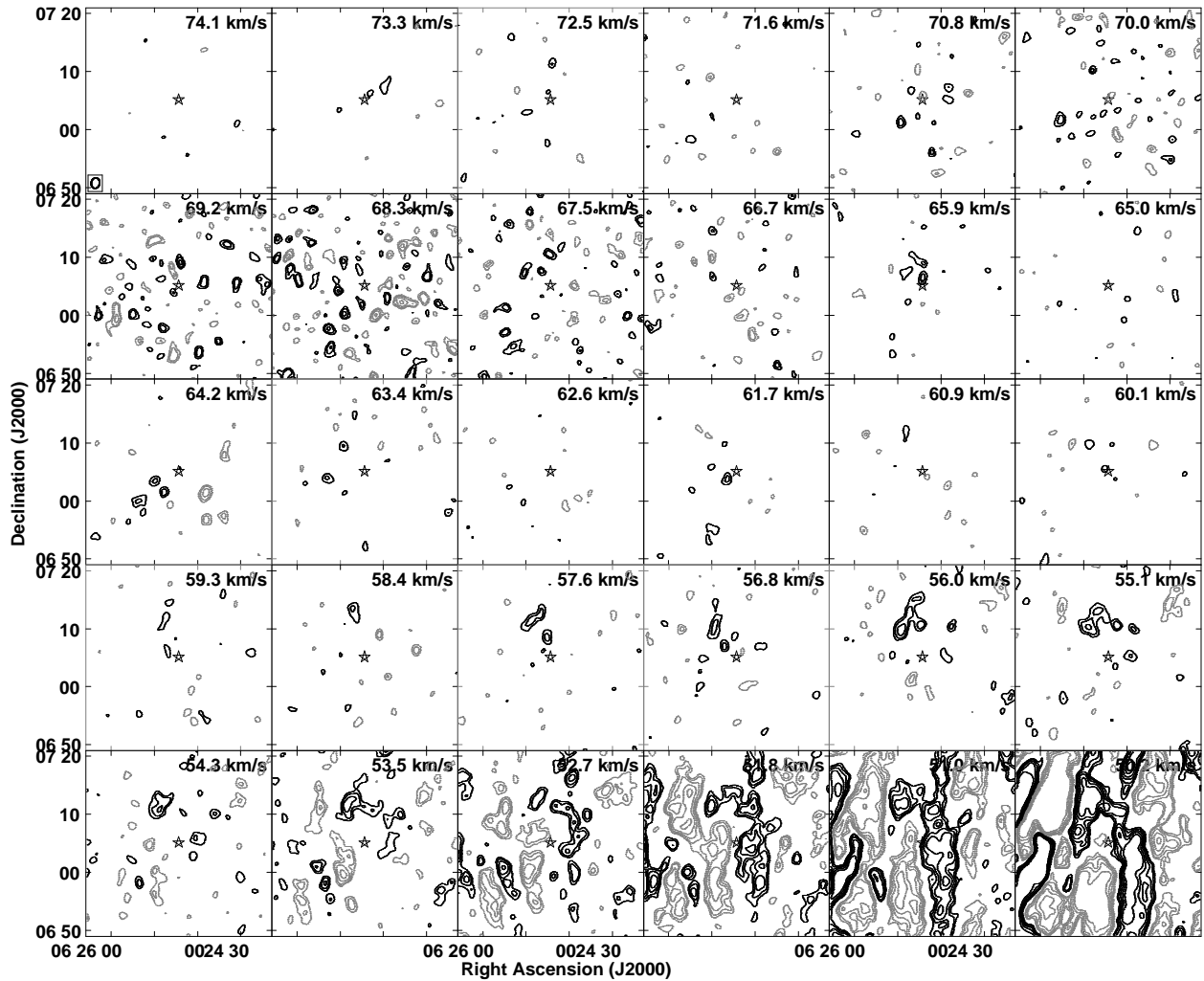}}}
\caption{As in Figure~\ref{fig:TMoncmapsblue}, but for velocities
  redshifted from the systemic velocity of T~Mon. Channels in the
  bottom row are dominated by large-scale Galactic emission along
  the line-of-sight. The elevated noise in channels from 66.7 to
  70.0~\kms\ results from aliasing at the subband boundaries in the
  WIDAR correlator.  }
\label{fig:TMoncmapsred}
\end{figure*}

\subsection{T Monocerotis (T Mon)\protect\label{TMon}}
Along the line-of-sight to T~Mon, the stellar systemic velocity
($V_{\star,{\rm LSR}}=16.1$~\kms) is coincident with strong interstellar
foreground/background emission. 
Consistent
with the LAB spectrum shown in 
Figure~\ref{fig:LABspec}, we find that the VLA \HI\ channel
images for T~Mon are affected by moderate to strong line-of-sight confusion 
over the velocity range $-13.3\lsim$ $V_{\rm LSR}$ $\lsim 64.2$~\kms. We
therefore focus our search for  circumstellar emission
outside of this window. 

In Figure~\ref{fig:TMoncmapsblue} we present \HI\ channel maps for a range
of velocities blueshifted by $\sim$30-54~\kms\ from the stellar
systemic velocity. Within the first few channels shown (which are closest
in velocity to the dominant Galactic emission), no obvious signs of
large-scale Galactic contamination are apparent, and the noise
is consistent with  thermal noise. In the bottom row of Figure~\ref{fig:TMoncmapsblue},
the elevated noise in several channels results from aliased noise at
the subband edges of the WIDAR correlator. 
However, in between, we find signatures of spatially extended
emission that are visible  at or near the stellar position  over several
contiguous channels with a
significance of $\ge3\sigma$.

% Fig. 4
% 
\begin{figure}
%\vspace{-1.5cm}
\centering
\scalebox{0.45}{\rotatebox{0}{\includegraphics{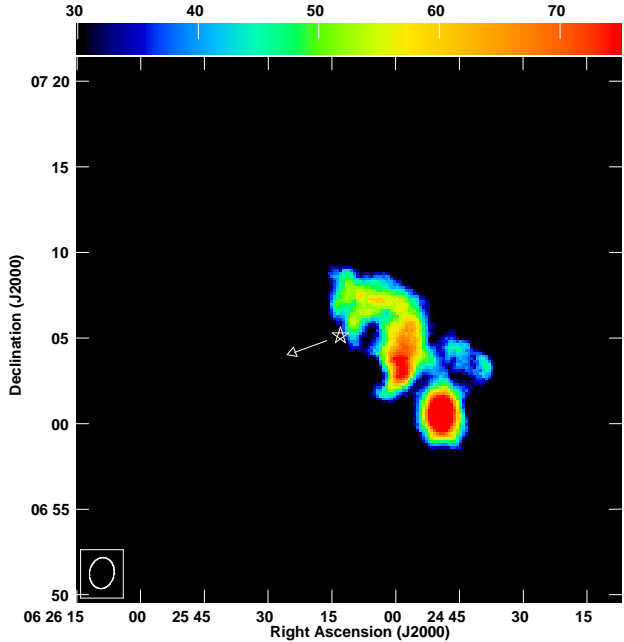}}}
\caption{\HI\ total intensity map of the region surrounding T~Mon,
  derived from a tapered data cube. The map incorporates emission over velocities $-33.9\le V_{\rm
  LSR}<-13.2$~\kms\ and $64.2< V_{\rm
  LSR}\le 65.9$~\kms. Spectral channels between $-13.2$ and 64.2~\kms\ were
  excluded owing to contamination from line-of-sight emission. A
cutoff of 1.8~mJy beam$^{-1}$ (1$\sigma$) was imposed after smoothing
the data with a Hanning function in frequency and a boxcar kernel
with a FWHM of 7 pixels ($70''$) spatially. Intensity levels are in
  units of Jy beam$^{-1}$m s$^{-1}$. The synthesized beam
  ($109''\times85''$) is
  shown in the lower left corner.  A star symbol indicates the
  position of T~Mon, and the arrow indicates the direction of space
  motion of the star.}
\label{fig:TMonmom0}
\end{figure}

% Fig. 5
% 
\begin{figure*}
%\vspace{-1.5cm}
\centering
\scalebox{0.7}{\rotatebox{90}{\includegraphics{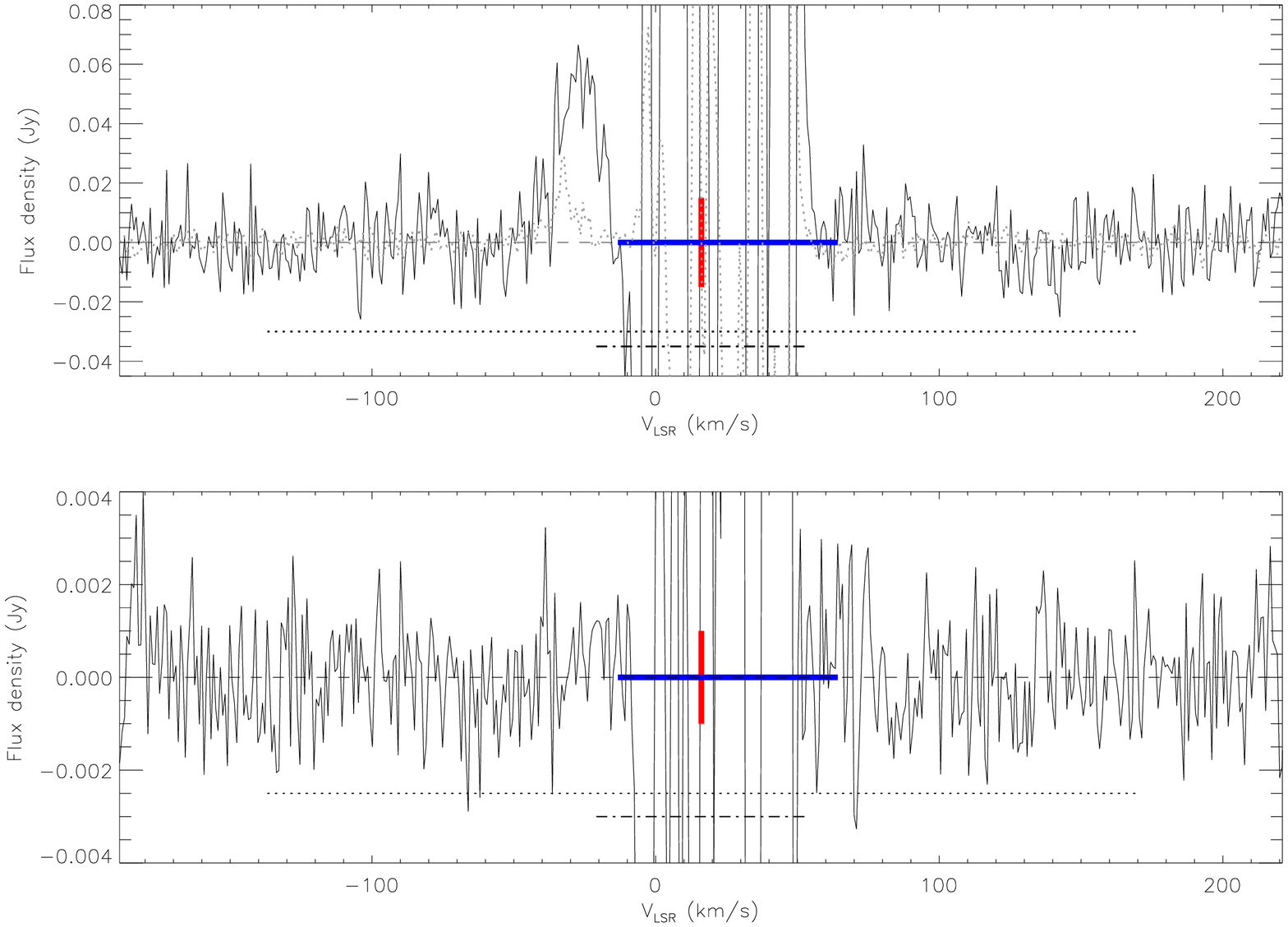}}}
\caption{VLA \HI\ spectra in the neighborhood of T~Mon, derived from
  naturally weighted data after correction for the primary beam. {\it
  Top:} The black solid line shows a spectrum integrated
  over a box with dimensions 
\am{6}{2}
east-west and \am{8}{2} north-south, centered on the emission nebula
  visible in Figure~\ref{fig:TMonmom0}. This aperture excludes the emission knot to the
  southwest.  A spectrum toward the emission knot itself is
  overplotted as a grey dotted line.   {\it Bottom:} Spectrum integrated over a single
  synthesized beam, centered on T~Mon. In both spectra, 
the red vertical lines indicate the stellar systemic
velocity and the blue horizontal lines denote the velocity ranges over
which analysis of the VLA data is impeded by 
contamination from interstellar emission along the
line-of-sight (cf. Figure~\ref{fig:LABspec}). 
The horizontal dotted lines indicate twice the escape velocity from the
stellar surface, and the horizontal dot-dash lines indicate the predicted
linewidth if the outflow obeys Reimers' (1977) formula (see
Section~\ref{upperlimits}). 
}
\label{fig:TMonHIspec}
\end{figure*}

Figure~\ref{fig:TMoncmapsred} shows a series of 
channels images 
redshifted from the stellar velocity. A 
few of these channels also show possible hints of extended emission
feature near the position
of the star, although they are weaker than the features seen
at comparable velocity offsets on the blue side of the stellar
systemic velocity, and in general
the identification of genuine features redward of $V_{\star, \rm sys}$
is hindered by aliased noise at
the subband edges (affecting velocities 66.7 to 70.8~\kms) and by line-of-sight
contamination (which affects velocities
$V_{\rm LSR}\lsim$64.2~\kms).

To further illustrate the nature of the emission visible in
Figure~\ref{fig:TMoncmapsblue} and \ref{fig:TMoncmapsred}, we show in
Figure~\ref{fig:TMonmom0} an \HI\ total intensity 
map derived from these data. 
This zeroth moment map was produced using emission between velocities $-33.9$
and 65.9~\kms. Channels between $-13.2$ and 64.2~\kms, which are
clearly dominated by Galactic contamination, were blanked, and a
cutoff of 1.8~mJy beam$^{-1}$ (1$\sigma$) was imposed after smoothing
the data with a Hanning function in frequency and a Gaussian kernel
with a FWHM of 7 pixels ($70''$) spatially.

Figure~\ref{fig:TMonmom0} reveals what appears to be a partial
shell-like structure several arcminutes across. At the distance of
T~Mon, the projected extent of the structure corresponds to $\sim$2~pc. The bulk of the
emission is offset to the northwest of the star, with only marginal
evidence of emission extending across the position of T~Mon itself.  However, it is
interesting to note that the offset between putative shell and T~Mon 
lies along the space trajectory of the star, indicated by an
arrow on Figure~\ref{fig:TMonmom0}. This space velocity vector was
computed using the 
distance and systemic velocity from Table~1 and the proper motion for T~Mon from
van~Leeuwen (2007); it corresponds to a 3D velocity of 21.0~\kms\
along a position angle of 110$^{\circ}$. Another noteworthy feature is that
in the direction trailing the arrow, roughly $2'$ behind the star (corresponding to a
projected distance of $\sim$0.82~pc), there appears to be
a depression or cavity in the \HI\ nebula.   Together these features raise the
intriguing possibility that the observed \HI\ emission corresponds to
material shed during an earlier epoch of mass loss from
T~Mon (see Section~\ref{TMondisc} for discussion).

Based on an examination of the channel maps in
Figure~\ref{fig:TMoncmapsblue},  the knot of
emission visible to the southwest of the putative shell (i.e., near
$\alpha_{\rm J2000}=06^{\rm h} 24^{\rm m}
49.6^{\rm s}$, $\delta_{\rm J2000}= 07^{\circ} 00'$ \as{41}{0}) appears to
be spatially and spectrally distinct from the remainder of the nebula
and is therefore likely to be unrelated. The line profile of this
knot peaks near $V_{\rm LSR}\approx-33.1$~\kms\ with a FWHM
of 3.7$\pm$2.2~\kms, and is
therefore blueshifted
relative to the bulk of the emission in the main nebula. Excluding this knot, the dimensions
of the remainder of the nebula are approximately \am{6}{1}$\times$\am{4}{5}
($\sim$2.6$\times$2.0~pc at the distance of T~Mon).

To illustrate the spectral characteristics of the emission in the
vicinity of T~Mon, 
we plot in the upper panel of Figure~\ref{fig:TMonHIspec} 
a spectrum integrated over a rectangular region encompassing the
nebula visible in Figure~\ref{fig:TMonmom0} with exception of the
southwestern knot. (A spectrum toward the knot is overplotted as a
grey dotted line for
comparison.) The aperture spanned \am{6}{2}
east-west and \am{8}{2} north-south, and
was centered at  $\alpha_{\rm J2000}=06^{\rm h}~25^{\rm m}~
05.6^{\rm s}$, $\delta_{\rm J2000}$=07$^{\circ}$ 04$'$ \as{58}{6}.

Outside of the velocity range dominated by Galactic
emission (designated by a blue line on  Figure~\ref{fig:TMonHIspec}),
we see evidence of 
an emission peak offset to the blue 
of the stellar systemic velocity by $\sim$45~\kms. There is also
marginal evidence for a much weaker redshifted peak at  $\sim$50~\kms\ from
$V_{\star,\rm sys}$.  

In the \HI\ total intensity image shown in Figure~\ref{fig:TMonHIspec},
\HI\ emission is only marginally detected at the position of T~Mon
itself. Consistent with this, a spectrum integrated over a single
synthesized beam centered on the star shows no significant emission
outside the spectral regions affected by confusion (lower panel of
Figure~\ref{fig:TMonHIspec}). 

To estimate the total quantity of gas associated with the nebula in
Figure~\ref{fig:TMonmom0}, we use the spectrum plotted 
in the upper panel of Figure~\ref{fig:TMonHIspec}.  
Integrating over the same velocity ranges used to derive the \HI\
total intensity map (and similarly excluding
the portion of the spectrum deemed contaminated by interstellar
emission), we find a velocity-integrated \HI\ flux density  of
$\int S dV\sim$0.76~Jy~\kms. Assuming the emission is optically thin, this
translates to an \HI\ mass of $M_{\rm HI}\approx0.36M_{\odot}$ at the distance of
the star (see 
Section~\ref{upperlimits}). 
This should be considered a lower limit to the total \HI\
mass given the range of velocities that was excluded because of confusion.
Applying a scaling factor of 1.4 to correct for the mass of helium,
this translates to a total nebula mass of $\gsim0.5M_{\odot}$. 

\subsection{$\zeta$ Geminorum ($\zeta$~Gem)\protect\label{zetaGem}}
As seen in Figure~\ref{fig:LABspec}, 
the line-of-sight \HI\ emission in the direction of
$\zeta$~Gem is much weaker and narrower in velocity extent
compared with the other three 
stars in the present sample. However,
despite the modest confusion, 
we do not identify any statistically significant line emission in
our \HI\ data cubes that
can be attributed to circumstellar gas.

\begin{figure*}
%\vspace{0.3cm}
\centering
\scalebox{0.7}{\rotatebox{90}{\includegraphics{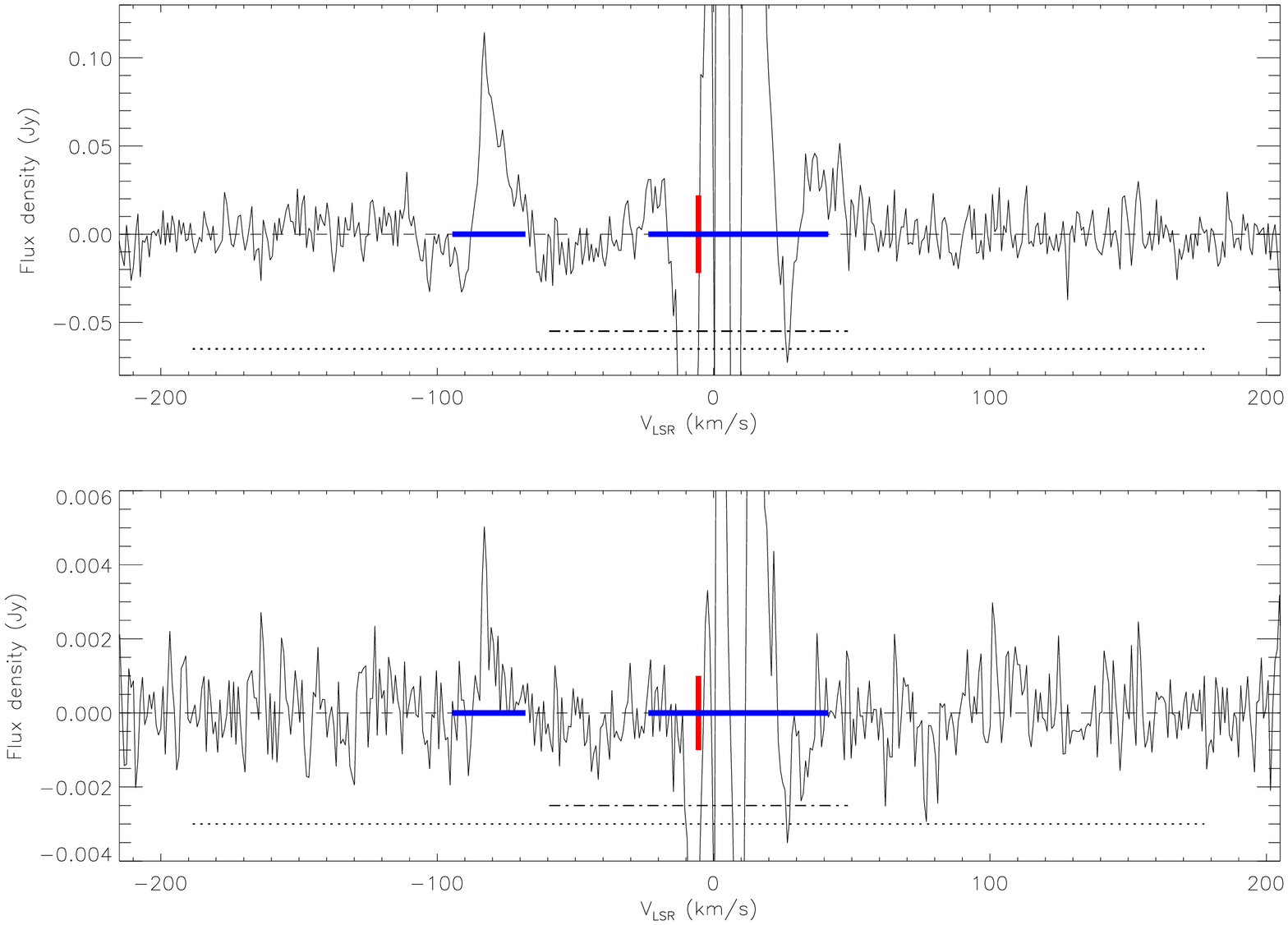}}}
\caption{{\it Top:} VLA \HI\ spectrum toward $\zeta$~Gem, derived
  from a naturally-weighted data cube by integrating
over a $\sim 1\times1$~pc box ($538''\times538''$)
surrounding $\zeta$~Gem. The data are corrected for the VLA primary beam.
{\it Bottom:} \HI\ spectrum integrated over a
single synthesized beam centered on the star. The meanings of the 
overplotted vertical and horizontal lines are as in Figure~\ref{fig:TMonHIspec}. }
\label{fig:zetaGemspec}
\end{figure*}

In Figure~\ref{fig:zetaGemspec} we plot two \HI\ spectra toward the
 position of $\zeta$~Gem derived from our VLA data. The top panel
 shows a spectrum integrated over a
 1~pc$^{2}$ box ($536''\times536''$) centered on the star, while the
 lower panel shows a spectrum averaged over a single synthesized beam
 at the stellar position. No significant spectral
 features are detected in either spectrum outside of the velocity range 
 that is dominated by line-of-sight confusion (indicated by  blue
 horizontal lines), with the possible exception of near velocities 
$\sim$40-45~\kms\ in the spatially integrated
 spectrum. However, the \HI\ channel maps (not shown) do not reveal any
corresponding emission regions with a significance of $>3\sigma$. 
 The  features in both spectra near
$-83$~\kms\ are clearly interstellar in nature; there is a counterpart  in the LAB data
(Figure~\ref{fig:LABspec}), and  the
 associated channel maps contain large-scale emission that
 fills the primary beam.  In Section~\ref{upperlimits} we discuss 
upper limits to the circumstellar \HI\
  mass and mass-loss rate of $\zeta$~Gem based on these results.

% Fig. 6
% 

\subsection{RS Puppis (RS Pup)\protect\label{RSPup}}
As noted in Section~\ref{targets}, Kervella et al. (2009) reported evidence
for a warm stellar wind from RS~Pup based on near-IR
interferometry observations. Our \HI\ 21-cm line observations 
now permit a search for a gaseous counterpart to this wind, as well as
for more
spatially extended circumstellar material to which near-IR interferometry
is insensitive. In addition,
our VLA data allow a search atomic hydrogen associated with the RS~Pup reflection
nebula. 
As discussed in Section~\ref{targets}, the extended reflection 
nebula surrounding RS~Pup is thought to comprise
predominantly swept-up interstellar matter rather than circumstellar
ejecta. Nonetheless, this nebula is believed to be physically associated
with the star, and its cold temperature ($\approx$40-45~K; 
McAlary \& Welch 1986; Kervella et al. 2009; Barmby
et al. 2011) make it a candidate for the presence of
associated \HI\ gas. 

The LAB spectrum shown 
in Figure~\ref{fig:LABspec} reveals that strong interstellar \HI\
emission is present over a wide range of velocities along the
direction toward RS~Pup. Correspondingly, our VLA channel images
are contaminated by line-of-sight emission over a
significant fraction of the observing band
($-25\lsim V_{\rm LSR}\lsim 171$~\kms).  This range encompasses the
stellar systemic velocity ($V_{\star,\rm LSR}=1.3$~\kms). 

\begin{figure*}
%\vspace{-0.3cm}
\centering
\scalebox{0.7}{\rotatebox{90}{\includegraphics{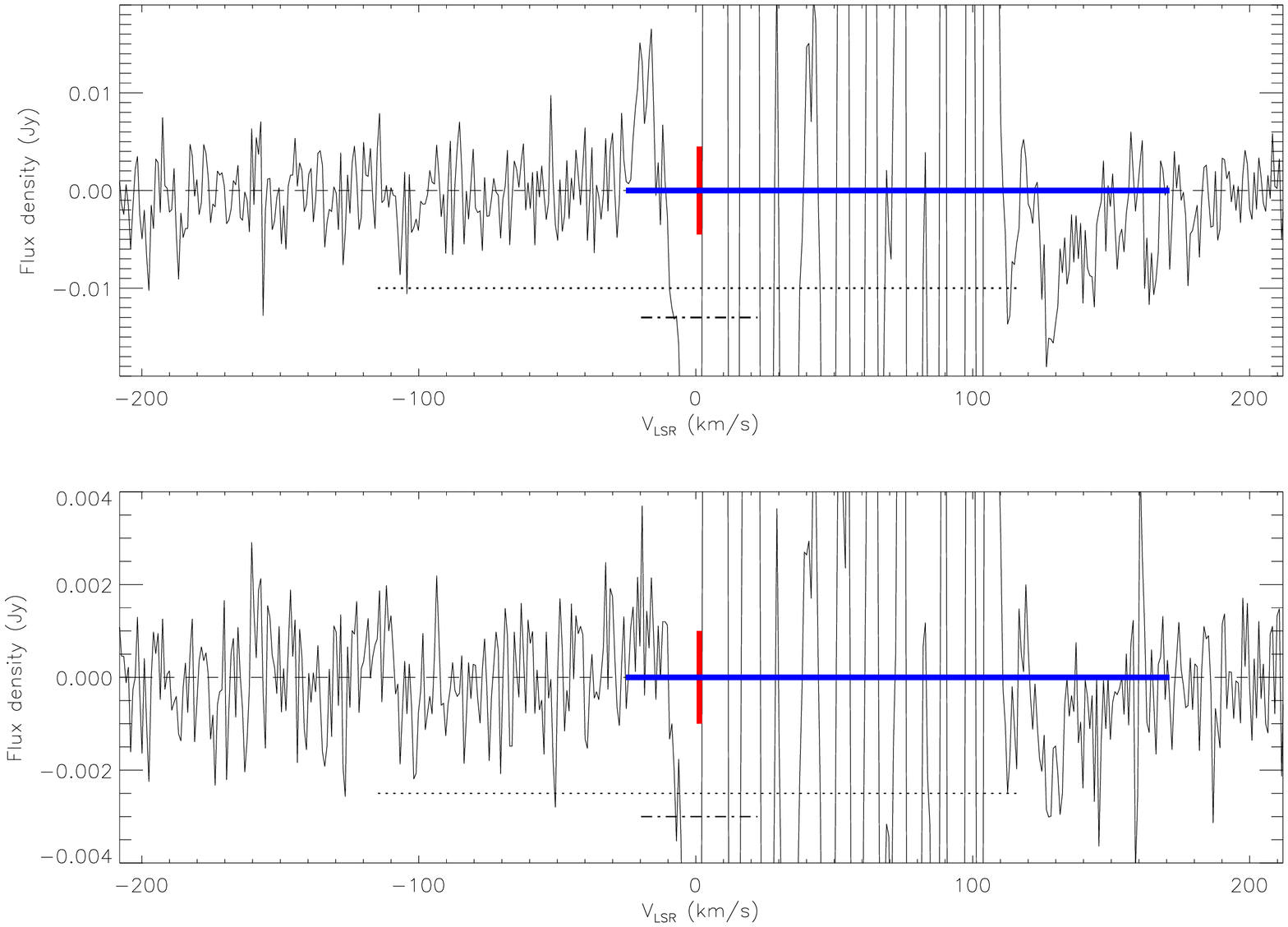}}}
\caption{As in Figure~\ref{fig:TMonHIspec}, but for RS~Pup. In the
  upper panel, the spectrum is integrated over a $136''\times136''$ 
($\sim 1\times1$~pc) box.   }
\label{fig:RSPupspec}
\end{figure*}

To illustrate the spectral characteristics of the \HI\ emission
near the position of RS~Pup,
we present in Figure~\ref{fig:RSPupspec} integrated \HI\ spectra
derived from our VLA data. The top panel shows  the emission integrated
over a 1~pc$^{2}$ ($136''\times136''$) box centered on the
star, while the lower panel shows a
spectrum integrated over a single synthesized beam centered on the
stellar position. The former region is roughly comparable in extent to the 70$\mu$m
emission nebula detected by {\it Spitzer} (cf. Barmby et al. 2011,
their Figure~5). 

Blueward of $V_{\rm LSR}\approx -25$~\kms, the Galactic emission
toward RS~Pup becomes
negligible, providing a clean window for identification of
possible circumstellar emission.  However, we find no statistically
significant emission near the position of RS~Pup either 
in our \HI\ channel maps (not shown) or in our integrated
spectra. Over the total angular extent of the reflection nebula
($\sim4'$; Kervella et al. 2012), spatial filtering of emission by the VLA
should not be a factor (see Section~\ref{observations}).
We conclude that any atomic hydrogen associated with mass loss
from RS~Pup or with the RS~Pup reflection nebula is either below our
detection threshold or lies at velocities affected by confusion from
line-of-sight emission.
In Section~\ref{upperlimits} we use these results to 
estimate upper limits to the circumstellar \HI\
  mass and mass-loss rate of RS~Pup.

% Fig. 7
% 

\subsection{X Cygnus (X Cyg)\protect\label{XCyg}}
\begin{figure*}
%\vspace{0.3cm}
\centering
\scalebox{0.7}{\rotatebox{90}{\includegraphics{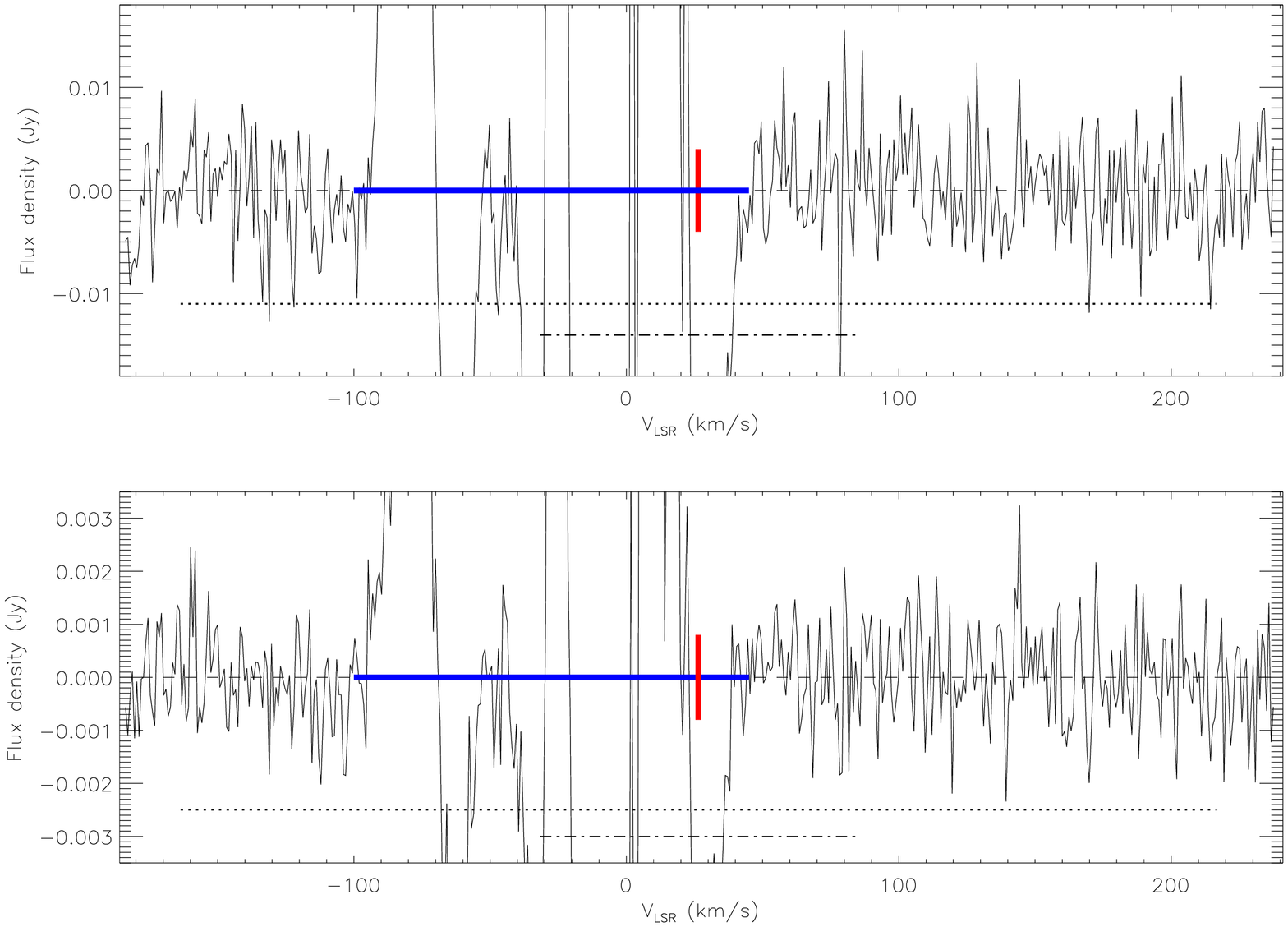}}}
\caption{ As in Figure~\ref{fig:TMonHIspec}, but for X~Cyg. In the
  upper panel, the spectrum is integrated over a $210''\times210''$ 
($\sim 1\times1$~pc) box.   }
\label{fig:XCygspec}
\end{figure*}

In the direction of X~Cyg, our VLA \HI\ channel maps over the
velocity range $-100\lsim$ $V_{\rm LSR}$ $\lsim 45$~\kms\ 
contain the hallmarks of contamination from line-of-sight emission. 
Consistent with this, the LAB spectrum toward the position
of X~Cyg (Figure~\ref{fig:LABspec}) shows the presence of
strong interstellar emission over this velocity range.  
In contrast, for velocities 
$V_{\rm LSR}\gsim$45~\kms, 
the LAB spectrum toward X~Cyg appears virtually free of contaminating
emission, providing a relatively clean spectral window to search for circumstellar
emission redshifted from the stellar systemic
velocity of $V_{\star,{\rm LSR}}=$26.4~\kms.
However, based on inspection of our \HI\ data cubes, we find no
statistically significant emission at or near the stellar position over
the velocity range $45\lsim V_{\rm LSR}\lsim 237$~\kms.
Similarly, spectra derived by summing the
emission over a 1~pc$^{2}$ ($210''\times210''$) box
surrounding the star (Figure~\ref{fig:XCygspec}, top) or a 
single synthesized beam centered on the star
(Figure~\ref{fig:XCygspec}, bottom) reveal no
signs of significant spectral features outside the velocity range
that is contaminated by line-of-sight confusion. We discuss upper limits on the
circumstellar \HI\ mass and mass-loss rate of X~Cyg in Section~\ref{upperlimits}.

% Fig. 8
% 

\section{T~Mon: An \HI\ Nebula Associated with
  Previous Mass Loss?\protect\label{TMondisc}}
\label{TMondisc}
While a chance superposition of an
unrelated gas cloud along the line-of-sight cannot be entirely
excluded as an explanation for the ``shell'' of emission seen near
T~Mon (Figure~\ref{fig:TMonmom0}), several
characteristics of this emission are consistent with a possible
association with the star. We now discuss these arguments in more depth, and consider the
implications for the evolutionary history of T~Mon if this shell is
indeed associated with stellar mass-loss.

\subsection{Implications of the Shell Morphology and Kinematics
on its Probable Origin}
It is now well-established that detectable circumstellar envelopes of atomic hydrogen,
spanning up to a parsec or more in size,
form around certain types of evolved, mass-losing stars, and
furthermore that the properties of these envelopes are significantly
impacted by the space motion of the stars and the interaction of the
stellar ejecta with the ISM (e.g., G\'erard \&
Le~Bertre 2006; Le~Bertre et al. 2012; Matthews et
al. 2013). For stars with low space velocities, quasi-spherical
shells may be observed, while in cases where the space velocity is sufficiently high,
bow shocks may be detectable in IR and ultraviolet wavelengths,
and the effects of ram pressure may sweep
back a portion of 
the ejecta  to form a trailing gaseous wake (see M12; Matthews et al. 2013 and references
therein). Numerical models also predict that the properties of the
shells will change significantly as stars evolve and undergo changes
in their mass-loss rates, and as the ejecta expand into the ISM 
(e.g., Wareing et al. 2007; Villaver et
al. 2012). However, 
one general consequence of the central star's space motion 
will be an eventual shift of the star from the geometric center of the shell to
a position nearer its leading edge (e.g., Smith 1976).
Once the mass loss wanes, the star will eventually completely exit
the nebula (Smith 1976; Wareing et al. 2007) and  leave only a partial shell
downstream (see Figure~7d of Wareing et al.). 

As already noted in Section~\ref{TMon} the space velocity vector of
T~Mon traces directly back to the approximate geometric center of the
detected \HI\ shell, consistent with the possibility that this shell
could be a mass-loss remnant. However, at the same
time, 
the fact that the star appears to be nearly completely outside
the shell's boundaries 
implies that if it is circumstellar in origin, the shell must have formed 
during a {\em previous} epoch of mass-loss that did not continue at
a sustained rate to the present day. The absence of detectable
emission along the leading edge of the nebula is also consistent with
it being a ``fossil'' shell (see Wareing et al. 2007).

The interaction between the ISM and stellar ejecta affects not only
the morphology of the circumstellar 
gas, but also its kinematics, including an appreciable
deceleration of the outflowing gas at large distances from the star (e.g.,
Matthews et al. 2008, 2011). Qualitatively, an excess of
blueshifted emission, as we see in the case of T~Mon, might therefore be
explained by the decelerating effect of the ISM on the ejecta (see
Matthews et al. 2008).  This would also imply that the red and blue peaks visible
in the spectrum in Figure~\ref{fig:TMonHIspec} cannot be directly interpreted
as a measure of the stellar outflow speed. 

Clearly numerical models that explore the interaction
of mass-losing Cepheids with their environment would be of
considerable interest for helping to interpret future \HI\
observations of Cepheids. 
Numerical models to date have focused on AGB and
post-AGB stars whose wind speeds, mass-loss rates, and mass-loss
timescales all may differ significantly from Cepheids. For example,
typical AGB stars have
outflow velocities  $V_{\rm o}\sim$5-10~\kms, which 
are generally several times smaller than those 
expected for Cepheids (see Section~\ref{upperlimits}). This means that AGB stars are
frequently in the regime where $V_{\rm 
o}$ is less than the space velocity $V_{\rm 3D}$, whereas Cepheids
will more typically have  $V_{\rm
  o}\gsim V_{\rm 3D}$ (cf. Table~1).

\subsection{Constraints on the Formation Time of a Possible T~Mon Shell}
As described above, the morphology of the \HI\ nebula northwest of
T~Mon appears to contain a depression or cavity along its southeast
edge. Possible 
explanations for such a cavity in a circumstellar shell could be
instabilities along the leading edge of a shell (see Matthews et
al. 2013) or the cessation of mass loss from the central star. 
Assuming the nebula lies at the same distance as T~Mon, 
tracing the location of the star backwards along its projected
space motion vector places it within this cavity at some time in the
past (assuming the nebula
does not have a transverse velocity component different from
that of the star). 
Given the space velocity of T~Mon (Table~1), an estimate of the travel time from the
center of the apparent \HI\ cavity to the current stellar position is
$\gsim$38,000 years, although this is a lower limit to the elapsed
time, both because of projection effects and 
because it assumes that the nebula has remained stationary in the local ISM rest
frame. 

T~Mon is believed to be currently on its third crossing of the
instability strip (Turner \& Berdnikov 2001), and based on Bono et al.
(2000), we estimate the total lifetime of the
third crossing to be 
$t_{c}\sim26,000$~yr for a star of the mass and period of T~Mon (assuming solar
metallicity).\footnote{The models of Bono et al. (2000) assume no
  mass loss.} This implies that if the \HI\ nebula comprises mass loss on
the instability strip, it must
have occurred during the {\em second} crossing, whose duration is estimated to
be $t_{c}\sim$53,000~yr. (The first crossing can be largely excluded
based on its short lifetime of $t_{c}\sim$3600~yr).

\subsection{Implications of Mass Loss on the Evolutionary History
  of T~Mon\protect\label{tmonimp}}
Based on Bono et al. (2001), the mass discrepancy for T~Mon is
$\sim0.9~M_{\odot}$ assuming models that do not include core
overshoot (i.e., using ``canonical masses''), while for 
non-canonical masses, the discrepancy would increase slightly to
$\sim0.95~M_{\odot}$. These values are of order 10\% of the mass of
the star.  Interestingly, this is within a factor of two
of the mass we estimate for the nebula  near the star ($\gsim
0.5M_{\odot}$; Section~\ref{TMon}).

Assuming this nebula indeed
represents a circumstellar remnant, some
fraction of its mass should have originated from material swept from the surrounding
ISM.  If we make a crude estimate of that contribution by assuming the distance from the Galactic Plane
is $z\approx d~{\rm sin}b +15$~pc and the local ISM number density is
approximately $n_{\rm H}(z)=2.0e^{-(|z|/100~{\rm pc})}$ (Loup et
al. 1993), we find $n_{\rm H}\approx$~1.2~cm$^{-3}$. Assuming a
geometric mean
  radius for the shell of 1.0~pc (see Section~\ref{TMon}), the measured atomic
  hydrogen mass therefore translates to a particle density $n_{\rm
    H}\approx$3.5~cm$^{-3}$, or roughly three times the expected local
  density. Thus it is plausible that as much as two-thirds of the shell mass
  ($\sim0.4~M_{\odot}$) may have had a mass-loss origin---an amount 
sufficient to reconcile a significant part of T~Mon's mass
discrepancy and to have an
important impact on the evolutionary history of the star. 

For a fiducial outflow speed of $\sim$40~\kms\ (Table~5; see also Section~\ref{upperlimits}), a
lower limit to the expansion age of the T~Mon nebula is $\sim$24,000~yr.
Assuming that two-thirds of the
mass of the nebula was shed from the star over a comparable
period would imply a mass-loss rate of ${\dot
  M}\lsim 2\times10^{-5}~M_{\odot}$~yr$^{-1}$.  
On the other hand, if we take as an upper limit to the age of the
nebula the time that T~Mon has spent on the instability strip
($\sim70,000$~yr; Bono et al. 2000), this would translate to ${\dot
M}\sim6\times10^{-6}~M_{\odot}$~yr$^{-1}$. 

The first of these two estimates is
two orders of magnitude higher than the mass-loss rate estimated by
Gallenne et al. (2013) based on the modeling of extended mid-IR
emission [${\dot
  M}\approx(5.6\pm0.6)\times10^{-7}~M_{\odot}$ yr$^{-1}$, after scaling to our
adopted distance]. However, given that the observations of Gallenne et
al. sample material within only $\sim10-20R_{\star}$, the mass-loss rate derived by
those authors reflects only very recent or ongoing mass loss. 
As described above, \HI\ emission is only marginally detected at the
position of T~Mon itself (see Figures~\ref{fig:TMonmom0} and
\ref{fig:TMonHIspec}, lower panel). 
Using the same approach as for the
three undetected stars in our sample (Section~\ref{upperlimits}), 
we may therefore place a 3$\sigma$ upper limit on
the {\it current} mass-loss rate for T~Mon of ${\dot
  M}<4.2\times10^{-5}~M_{\odot}$~yr$^{-1}$. Our \HI\
observations are therefore not sensitive to ongoing mass loss at a rate
comparable to that estimated by
Gallenne et al. 

Regardless of the current mass-loss rate of T~Mon, the location of the
star relative to the \HI\ nebula implies that if the nebula is
circumstellar in origin, then the star must have ceased
losing mass for some period of time since the nebula's
formation. Combined, the new \HI\ data and 
previous mid-IR
results of Gallenne et al. (2013) point to a scenario where T~Mon may have undergone
significant variations in its mass-loss rate over the past 25,000 years
or more. This is consistent with  theoretical predictions that such
fluctuations will occur, particularly in long-period Cepheids (e.g., Neilson et
al. 2011).  Indeed, as we have noted above, this disruption could have
resulted from the star temporarily leaving the instability strip after its second
crossing.  

Finally, it is also worth noting that
T~Mon, as well as $\delta$~Cep, whose circumstellar material was
detected in \HI\ by M12, are both members of triple systems. 
The presence of  a close companion (as in the case of $\delta$~Cep) 
may affect the nature and intensity of
mass-loss episodes. However, even companions at larger separations
(present in both cases) may have
an important effect on the physical
properties 
of stellar ejecta (Mohamed \& Podsiadlowski 2012; 
Anderson et al. 2015). While it is not possible to
draw general conclusions based on a sample of only two stars,
further theoretical modeling, in addition to observations
of a larger sample of binary Cepheids in \HI\ and other wavelengths
sensitive to mass loss, could provide additional
insights into the consequences of binarity.

\section{Discussion}
\subsection{Upper Limits on the Circumstellar \HI\ Mass and Mass-Loss
  Rates for the Undetected Stars\protect\label{upperlimits}}
For the stars undetected in the \HI\ line, our new
VLA observations allow us to place new limits on the presence of
circumstellar debris that may
have been shed by these Cepheids during periods of recent or ongoing mass loss.
However, 
translating our measurements to quantitative
upper limits that are useful for constraining
the stellar mass-loss properties requires adopting some assumptions about the nature of
Cepheid outflows---a topic on which we have very few
empirical constraints.  

To our knowledge, $\delta$~Cep is only Cepheid with a
directly measured outflow velocity ($V_{\rm o}\approx$35~\kms; M12).  The value of $V_{\rm o}$ for
$\delta$~Cep is noteworthy
in that it is
significantly lower than the escape speed from the star, consistent with a
general trend of $V_{\rm o}<<V_{\rm esc}$ that has been seen in
other types of supergiants (Reimers 1977; Holzer \& MacGregor
1985; Judge \& Stencel 1991).
Indeed, Reimers (1977) found that for non-variable
G and K supergiants, the stars roughly follow a relation of the form $V_{\rm
  o}\sim 1.6\times10^{-3} V_{\rm esc}^{2}$. This relationship also holds for stars
of comparable spectral type in the sample of Judge \& Stencel (1991)
and reasonably agrees with the measured outflow speed for $\delta$~Cep
(M12), even though the underlying mass-loss mechanism may be quite
different between pulsating and non-pulsating supergiants.
Lacking any further constraints on $V_{\rm o}$ for
Cepheids, we adopt the empirical relation of Reimers to
estimate representative outflow velocities for our sample
stars. These values are
presented in Table~5. 

If we assume that the \HI\ linewidth for each star is
approximately twice its outflow speed,  we
may now derive model-dependent limits on the quantity of
circumstellar gas for the undetected stars. For $\zeta$~Gem, X~Cyg,
and  T~Mon, we use
the RMS noise levels, $\sigma_{\rm RMS}$, from the uncontaminated
portions of the
naturally-weighted \HI\ data cubes in Table~4 to compute
3$\sigma$ upper limits to the velocity-integrated
\HI\ flux density within a single synthesized beam centered on each of the
undetected stars as $\int S dV < 3\sigma_{\rm RMS}\times (2V_{\rm
  o})$ Jy \kms. For RS~Pup, where the entire velocity range
$V_{\star,\rm LSR}\pm V_{\rm o}$ is affected by confusion (see
Figure~\ref{fig:RSPupspec}), we substitute for $\sigma_{\rm RMS}$ the term
$\sigma_{\rm obs}=(\sigma^{2}_{\rm c}+\sigma^{2}_{\rm RMS})^{0.5}$, where $\sigma_{c}$=0.66~mJy
beam$^{-1}$ is the additional confusion noise estimated from channels 
with velocities between $-25$ and
$-10$~\kms.  
For optically thin emission, the aforementioned upper limits
to the integrated flux density can be
translated into 3$\sigma$ upper limits on the mass of \HI\ within the
synthesized beam as
$M_{\rm HI}<2.36\times10^{-7}d^{2}\int S dV~~M_{\odot}$, where $d$ is the adopted
distance in pc (e.g., Roberts 1975).  Results are given in column 4 of
Table~5.

Because the mass loss on the instability strip is
expected to extend over
tens of thousands of years or more, ejecta may be spread well beyond 
a single beam diameter---possible reaching a parsec or more from the
star (see M12; Kervella et al. 2012; Section~\ref{TMon}). 
For each undetected star, we therefore also compute
upper limits on the total \HI\ mass within a fiducial volume of radius
0.5~pc. The choice of this radius is arbitrary, but is useful for
illustrative purposes. These resulting limits are given in column~7 of Table~5. 

To provide an estimate of the rate of recent or ongoing
mass-loss for each star, we assume ${\dot M}< 1.4(M_{\rm HI}/t_{c})$. 
Here, the factor of 1.4 accounts for the mass of helium, 
and the fiducial timescale $t_{\rm c}$ is taken as 
$r_{c}/V_{o}$. We adopt as  the characteristic radius, $r_{c}$, 
the geometric mean
HWHM of the synthesized beam, projected to the distance of the star. 
Results are given in column~5 of Table~5. We include an upper limit
for T~Mon, where \HI\ was detected offset from the stellar position,
but not directly along the line-of-sight to the star (see Section~\ref{TMondisc}).

To constrain the {\em mean} mass-loss rates of $\zeta$~Gem, RS~Pup,
and X~Cyg during their entire Cepheid evolution, 
we combine the volume-averaged \HI\ mass limits computed above
with estimates of
the total time, $t_{i}$, that each of the stars have spent on the
instability strip. We estimate $t_{i}$ 
based on the solar metallicity models of Bono et
al. (2000, their Table~7). We assume that
$\zeta$~Gem is on its second crossing of the instability strip, that RS~Pup and X~Cyg are on
their third crossings, and that the total time spent on the
instability strip is equal to the sum of the previous crossings, plus one-half
the predicted duration of the current crossing. The resulting
time-averaged mass-loss rates, $|{\dot M}|=1.4\left(M_{\rm HI}({\rm
  total})\right)/t_{i}$, 
are given in column~8 of Table~5.  

\subsection{Upper Limits Compared with Expected Values of
  ${\dot M}$}
Based on recent studies of rates of period change, 
the mean mass-loss rate expected over the course of a Cepheid's
lifetime is
$\sim10^{-7}$ to $10^{-6}M_{\odot}$ yr$^{-1}$ (Neilson et
al. 2011, 2012a, b).   These values are comparable to the upper limits
in column~8 of Table~5. The present non-detection of $\zeta$~Gem,
X~Cyg, and RS~Pup in the \HI\ line is
therefore not in contradiction with the findings from the period change
studies, and suggests that deeper \HI\ observations may yet uncover mass-loss
signatures.  
Furthermore, 
the mass-loss rate of Cepheids are not expected to be constant, but
rather may vary  by up to several orders of magnitude as the stars evolve along the
instability strip 
(Neilson et al. 2011). This means that periods of intense mass loss 
($\sim10^{-5}M_{\odot}$ yr$^{-1}$)
may occur, particularly for longer period Cepheids ($P>$15~days;
B\"ohm-Vitense \& Love 1994; Deasy 1988; Neilson \& Lester
2008).  While the mechanism for generating such intense mass loss is unclear,  it is
worth noting 
that because the crossing time of the instability strip is
relatively short for Cepheids with periods of $\sim$15-30 days ($<10^{5}$~yr; Bono et al. 2000) 
periods of intense mass-loss are likely to be required if stars in
this period range (including X~Cyg
and T~Mon) are to lose even a
few per cent of their mass during the Cepheid phase (see
Section~\ref{discrep}).   
For $\zeta$~Gem and
RS~Pup, our present upper limits on the current mass-loss rates are
inconsistent with
mass loss of this magnitude during the past several thousand
years,  but for X~Cyg or T~Mon it cannot be excluded (see also
Section~\ref{TMondisc}). 

\subsection{Constraints on the Role of Mass Loss for Solving 
the Mass Discrepancy\protect\label{discrep}}
After scaling the upper limits to the circumstellar \HI\
mass within a 0.5~pc radius around each star (column~7 of Table~5)
by a factor of 1.4 to correct for the mass of He, it is of interest to
compare the resulting masses with the 
stellar masses from Table~1. We find that for the three undetected
stars, our limits on the mass
of circumstellar matter correspond to $\lsim$2-5\% of the stellar
mass.  

While
it is difficult to accurately estimate the mass discrepancy  for any
individual star owing to model uncertainties, statistically, discrepancies
between pulsation and evolutionary masses
average between 10-20\%  
(see Section~\ref{intro}). This tentatively
suggests that for our sample stars, mass loss alone is unlikely to
fully reconcile the mass
discrepancy, although it could still account for a significant fraction
of it. However, we stress that this conclusion is model-dependent. 
For example, if the \HI\ linewidths are
smaller than we have assumed (e.g., as a result of deceleration of
large-scale ejecta owing to interaction with the
surrounding ISM), this
could allow significant quantities of gas to be hidden by
line-of-sight confusion (see e.g., Le~Bertre et al. 2012). Alternatively,
if we have systematically underestimated the
expected wind outflow velocities, the inferred upper limits would also increase.

\subsection{Comparison to Past Results}
A comparison between our new results and previous constraints on the
mass-loss rate of T~Mon were described in Section~\ref{tmonimp}. Here we briefly 
compare our new results for the other three stars to earlier studies.

\subsubsection{$\zeta$ Gem} For 
$\zeta$~Gem, Sasselov \& Lester (1994) reported evidence based on
the \HeI~$\lambda$10830 line for the
outflow of material in the upper chromosphere, albeit with velocities
well below the escape speed (they found the mean \HeI\ line velocity
to be blueshifted from the stellar
velocity by 31~\kms).
Ultraviolet spectroscopy by Schmidt \& Parsons
(1984) and Deasy (1988) also
revealed possible outflow signatures in the Mg~II~h and k
line profiles of $\zeta$~Gem (see also Deasy \& Wayman 1986).  In this case, two
blueshifted components are seen with velocities
comparable to the surface escape velocity 
($\gsim-110$~\kms\ relative to the stellar systemic velocity). While
it is unclear whether  the large Mg~II~h and k velocities
are reflective of the bulk
outflow speed, 
as described in Section~\ref{zetaGem}, we find no statistically significant emission at comparable
velocities in our \HI\ data. In any case, a wind resulting from mass-loss at a
rate comparable to that estimated by Deasy (${\dot M}\sim10^{-10}~M_{\odot}$ yr$^{-1}$)
would be several orders of magnitude below the detection limit of our
\HI\ observations, although it is important to stress that 
Deasy's ${\dot M}$ value is a lower limit, since it
does not take into account the continuous flow of matter from the
upper atmosphere. 

\subsubsection{RS Pup} The recombination line study of Gallenne et al. (2011)
provided evidence of a significant quantity of atomic hydrogen in
the close environment of RS~Pup (i.e., on scales of $\sim1''$ or $\sim$1550~AU).
 Although our current spatial resolution is comparatively 
coarse, we are able to place a
3$\sigma$ upper limit on the mass of neutral atomic hydrogen within a radius of
38,000~AU from the star (i.e., one synthesized beam) of $<0.11~M_{\rm HI}$
(Table~5).

Looking to larger scales,
Kervella et al. (2012) found a mean radius of the RS~Pup reflection
nebula to be \am{1}{8} ($\sim$0.8~pc for our adopted distance) based on the 
analysis of scattered light images, and they estimated the total
quantity of gas plus dust within this volume to be $190~M_{\odot}$ (with an
uncertainty of $\sim$40\%). The assumed dust fraction is 1\%. Despite
the significant  line-of-sight contamination in our RS~Pup
data, such a large quantity of gas within a region spanning only a few
arcminutes in spatial extent should have been readily detectable
($>5\sigma$) in our data at
velocities blueshifted by $\gsim-10$~\kms\ from the stellar systemic
velocity, even if it were only $\sim$1\% atomic. 
This suggests that either
the nebula is predominantly molecular---consistent with the  
mean nucleon density of
$\sim$2600 cm$^{-3}$ implied from the work of Kervella et al. (2012)--- 
or else that the atomic gas associated with the reflection nebula lies
within the 
range of velocities where detection is hampered by line-of-sight
contamination (cf. Figure~\ref{fig:RSPupspec}).   
Lastly, it is worth noting that several previous authors derived much
more modest mass estimates for the nebula based on dust measurements in the IR 
[e.g., $\sim2.3M_{\odot}$ (McAlary \& Welch 1986); 
$\sim2.2M_{\odot}$ (Deasy 1988); $\sim$0.06 to $0.9M_{\odot}$; (Barmby
  et al. 2011)].\footnote{All of these estimates assume a gas-to-dust ratio of 100 and are
scaled to our adopted distance.} However, in contrast to these other
studies, the technique of measuring
scattered light used by
Kervella et al. probes additional dust content whose temperature
is too low to directly emit in the IR.

Based on {\it IRAS} data, 
Deasy (1988) previously estimated the mean rate of mass-loss from RS~Pup to be
$\sim3\times10^{-6}~M_{\odot}$
yr$^{-1}$ (scaled to our adopted distance). However, based on the structure of the surrounding nebula, 
he argued that the mass-loss from this star is likely to
be intermittent, with episodes of enhanced mass loss at rates as high
as a few
times $10^{-5}~M_{\odot}$ yr$^{-1}$.  Our present upper limits on the
mass-loss rate of RS~Pup (Table~5) appear to
exclude ongoing mass loss of this magnitude. 

\subsubsection{X Cyg} Based on {\it IRAS} data, McAlary \& Welch (1986) noted a possible
IR excess associated with X~Cyg. Barmby et al. (2011) also found
tentative evidence for extended IR emission around this star in their {\it
  Spitzer} images. However,
in neither case are the data sufficient to estimate a mass-loss rate,
and to our knowledge,  no empirical limits on the mass-loss rate of X~Cyg have been
published to date.

%%%%%%%%%%%%%%%%%%%%%%%%%%%%%%%%%%%%%%%%%%%%%%%%%%%%%%%%%%%%%%%%%%%
% --- Table 5: Results
%
\begin{deluxetable*}{lcccccccc}
\tabletypesize{\tiny}
\tablewidth{0pc}
\tablenum{5}
\tablecaption{Derived Mass-Loss Properties for the Target Stars}
\tablehead{
\colhead{Star} & \colhead{$V_{\rm esc}$} & \colhead{$V_{\rm o}$} & 
\colhead{$\int S dv$} & \colhead{$M_{\rm HI}$(beam)}  &
\colhead{${\dot M}$(current)}  & \colhead{$M_{\rm
    HI}$(total)} & \colhead{$t_{i}$} &
\colhead{$|{\dot M}|$}\\
\colhead{}     & \colhead{(\kms)} & \colhead{(\kms)} & 
  \colhead{(Jy km s$^{-1}$)} & \colhead{($M_{\odot}$)} 
& \colhead{($M_{\odot}$ yr$^{-1}$)} & \colhead{($M_{\odot}$)} &
  \colhead{($10^{4}$ yr)} & \colhead{($M_{\odot}$ yr$^{-1}$)}\\
\colhead{(1)} & \colhead{(2)} & \colhead{(3)} & \colhead{(4)} &
\colhead{(5)} & \colhead{(6)} & \colhead{(7)} & \colhead{(8)} & \colhead{(9)} }

\startdata

$\zeta$~Gem & 183  & 54 & $<$0.35 & $<$0.012 & $<$1.9$\times10^{-5}$ &
$<$0.102 & 4.0 & $<2.0\times10^{-6}$\\

RS Pup& 116 & 21 & $<$0.20  & $<$0.112 & $<1.9\times10^{-5}$  & $<$0.258 &
140. & $<2.6\times10^{-7}$\\

X~Cyg & 190 & 58 & $<$0.41 & $<$0.093 & $<$5.6$\times10^{-5}$  & $<$0.288
& 7.2 & $<5.6\times10^{-6}$ \\

T~Mon$^{*}$ &153  & 37 & $<$0.33 & $<$0.156 & $<4.2\times10^{-5}$ &
0.4 & 7.0 & $\sim6\times10^{-6}$\\

\enddata

\tablenotetext{*}{For T~Mon, upper limits in columns 4-6 apply to 
  {\em recent or ongoing} mass-loss, not to a period of mass loss associated with
  the formation of the nebula northeast of the star (see
  Sections~\ref{TMon} and \ref{TMondisc}).}
\tablecomments{Explanation of columns: (1) star name; (2) escape
  velocity, assuming the stellar mass and radius from Table~1; (3)
  predicted wind outflow speed based on the relation $V_{\rm o}\approx
  1.6\times10^{-3} V_{\rm esc}^{2}$ (Reimers 1977); (4) 3$\sigma$
  upper limit on the integrated \HI\ flux density within a single
  synthesized beam centered on the star, assuming a rectangular line
  profile with a total linewidth of twice the estimated outflow velocity; (5) 
3$\sigma$ upper limit on the \HI\ mass within a single
  synthesized beam centered on the star, using the integrated flux density from
  Column~4; (6)  3$\sigma$ upper limit on the current stellar mass-loss rate
  (see Section~\ref{upperlimits} and Section~\ref{TMondisc} for
  details); (7) total detected \HI\ mass (for T~Mon) or 3$\sigma$ upper limit on the \HI\ mass within a
  volume of radius 0.5~pc surrounding the star; (8) estimated time spent on the
  instability strip based on Bono et al. (2000; see text for details); 
(9) estimated mass-loss rate (or 3$\sigma$ upper limit),
  averaged over the lifetime of the star on the instability
  strip.  }

\end{deluxetable*}

%%%%%%%%%%%%%%%%%%%%%%%%%%%%%%%%%%%%%%%%%%%%%%%%%%%%%%%%%%%%%%%%%%%

\section{Summary} 
We have presented \HI\ 21-cm line observations for a sample of
four Galactic Cepheids. Our goal was to search for circumstellar
gas associated with previous or ongoing mass loss. 
If present, such matter would help to reconcile the persistent
discrepancies of $\sim$10-20\% between the masses of Cepheids derived
from stellar evolution models versus those from 
the mass-dependent period-luminosity relation or orbital dynamics.
  
We have discovered a shell-like structure near the
long-period binary Cepheid, T~Mon.  
The star lies in projection just outside the edge of this
structure, with an offset from the geometric center in the direction of
the star's space motion. At the distance of T~Mon, the shell would have an atomic
hydrogen mass of $\sim0.4M_{\odot}$ and a size of
$\sim$2~pc. Although we cannot strictly exclude that the
shell was formed during a red supergiant phase, or
alternatively, that it represents a chance superposition of an
interstellar cloud along the
line-of-sight, its properties appear to be consistent with 
a fossil circumstellar shell that resulted from an earlier epoch of
mass-loss during T~Mon's previous crossing of the
instability strip. This interpretation would support a model where
mass-loss on the Cepheid instability strip is sporadic. Assuming that
approximately two-thirds of the shell's mass
originated from a stellar outflow, the mass of the material would be
sufficient to account for $\sim$50\% of the discrepancy between the
pulsation and evolutionary mass of T~Mon. 

For the other three stars in our sample (RS~Pup, X~Cyg, and
$\zeta$~Gem), no \HI\ emission was detected that
could be unambiguously associated with the circumstellar environment.
In all
three cases, line-of-sight confusion precluded searches of portions of
the observing band, with the effect being most severe for RS~Pup. For
the undetected stars, we place model-dependent 3$\sigma$ upper
limits on the  mass of circumstellar gas within a volume of radius
0.5~pc surrounding each star.  The resulting upper limits correspond
to $\lsim$2-5\% of the
respective stellar masses. Given typical Cepheid mass discrepancies of
$\sim$10-20\%, mass-loss through a neutral
atomic wind therefore cannot yet be excluded as making a significant contribution to
reconciling the discrepancy between pulsation and evolutionary mass
for these stars. Furthermore, our
upper limits on the mean mass-loss rates over their lifetimes on the instability strip
[$\lsim(0.3-6)\times10^{-6}~M_{\odot}$ yr$^{-1}$] are consistent with
mean mass loss rates derived from previous studies of the rate of
period change of large samples of Cepheids. 
However, we emphasize that our current upper limits are dependent on
uncertain assumptions about the properties of Cepheid winds, including the
predicted outflow velocity (see Section~\ref{upperlimits}). 

The findings here, together  with the previous work of M12,  suggest
that a deep \HI\ survey of a larger sample of nearby Galactic  Cepheids ($d\lsim$1~kpc)
could provide additional new constraints on the role of mass loss during the
Cepheid evolutionary phase. An order of magnitude increase in
sensitivity, such as could be achieved with next-generation
interferometers, 
would be particularly valuable. Additionally, the impact of confusion
could likely
be reduced in future observations through inclusion of data from longer baselines  
(e.g., Le~Bertre et al. 2012)
to help filter out the large-scale line-of-sight
contamination.

\acknowledgements
LDM is supported by
grant AST-1310930 from the National Science Foundation. Support to NRE
was provided from the {\it Chandra} X-Ray Center NASA contract NAS8-03060.
The observations presented here were part of NRAO program AM1087 (VLA/11B-035).
This research has made use of the SIMBAD database,
operated at CDS, Strasbourg, France.

%%%%%%%%%%%%%%%%%%%%%%%%%%%%%%%%%%%%%%%%%%%%%%%%%%%%%%%%%%%%%%%%%%%%%%%%%%

\end{document}